\let\AdSKernelLabel\label
\IfFileExists{revtex4.cls}{%
  \documentclass[aps,11pt,superscriptaddress,raggedbottom]{revtex4}
}{%
  \documentclass[aps,prd,11pt,superscriptaddress,raggedbottom]{revtex4-2}
}
\makeatletter
\@ifclassloaded{revtex4-2}{%
  \let\auto@bib\@empty
  \AtBeginDocument{%
    \def\frontmatter@abstractwidth{\textwidth}
    \tolerance=1000 \emergencystretch=4em
    \enlargethispage{-2\baselineskip}%
  }
}{}
\makeatother
\makeatletter
\@ifl@t@r\fmtversion{2023-06-01}{\let\label\AdSKernelLabel}{}
\makeatother
\usepackage{dcolumn,graphicx,amsmath,amsfonts,amssymb,bm}
\usepackage{color,cases,multirow,float,array,url,natbib,booktabs}
\usepackage{placeins}
\usepackage{subfigure}
\usepackage{hyperref}

\hypersetup{colorlinks=true,linkcolor=blue,citecolor=blue,urlcolor=blue,hypertexnames=false,pdftitle={Type-II quasinormal modes of dilaton--Euler--Heisenberg AdS black holes},pdfauthor={Chao-Ming Zhang, Fu-Wen Shu, Yun Soo Myung, Ming Zhang, De-Cheng Zou}}
\newcommand{\I}{\mathrm{i}}

\makeatletter
\g@addto@macro\normalsize{%
  \setlength{\abovedisplayskip}{6pt plus 1pt minus 1pt}%
  \setlength{\belowdisplayskip}{6pt plus 1pt minus 1pt}%
  \setlength{\abovedisplayshortskip}{3pt plus 1pt minus 1pt}%
  \setlength{\belowdisplayshortskip}{4pt plus 1pt minus 1pt}%
}
\makeatother

\begin{document}
\title{\resizebox{\textwidth}{!}{Type-II quasinormal modes of dilaton--Euler--Heisenberg AdS black holes}}
\author{Chao-Ming Zhang}
\email{chaomingzhang70@gmail.com}
\affiliation{Department of Physics, Nanchang University, Nanchang, 330031, China}

\author{Fu-Wen Shu}
\email[Corresponding author: ]{shufuwen@ncu.edu.cn}
\affiliation{Department of Physics, Nanchang University, Nanchang, 330031, China}

\author{Yun Soo Myung}
\email{ysmyung@inje.ac.kr}
\affiliation{Center for Quantum Spacetime, Sogang University, Seoul 04107, Republic of Korea}

\author{Ming Zhang}
\email{mingzhang0807@126.com}
\affiliation{Faculty of Science, Xihang University, Xi'an 710077, China}

\author{De-Cheng Zou}
\email{dczou@jxnu.edu.cn}
\affiliation{College of Physics and Communication Electronics, Jiangxi Normal University, Nanchang 330022, China}

\date[]{}
\begin{abstract}

 In this paper, we investigate  Type-II quasinormal modes (QNMs) of a magnetically
charged dilaton--Euler--Heisenberg black hole in anti-de Sitter (AdS)
spacetime. The Type-II includes  coupled gravitational, dilaton and
electromagnetic perturbations.
Its  coupled equations  are derived for $\ell=0,1,2$ modes and the QNM
frequencies are calculated by employing  direct integration and Chebyshev
pseudospectral methods, giving consistent results.
For nondegenerate dipole ($\ell=1$)  mode at fixed curvature, the nonlinear
correction starts at order $Q^2$ for
electromagnetic-led branch and at order $Q^4$ for scalar-led
branch. In this case,  stronger AdS curvature enhances reduction of the
electromagnetic-led real frequency for positive coupling, while
gravitational-led quadrupole real frequency changes weakly
with charge.
As the curvature weakens, the dipole frequencies approach the
AdS normal-mode scale and their damping decreases.
We find that all modes computed  have negative imaginary parts
for nonextremal black holes, implying this black hole is stable against Type-II perturbations.
Furthermore,  we explore  the connection between thermodynamic phase transition and QNMs for $\ell=0$ dilaton. 
For a subcritical isobaric process, the QNM real frequency of $\ell=0$ dilaton  decreases with increasing temperature of small black holes, while it increases with increasing temperature of large black holes. At coexistence, the damping rate for large black holes is about 4.28 times that for small black holes.

\end{abstract}
\maketitle

\section{Introduction}
\label{intro}

Quasinormal modes (QNMs) describe the damped oscillations of perturbed
black holes, with complex frequencies determined by the background
geometry, the dynamical fields and the boundary
conditions~\cite{Kokkotas:1999bd,Nollert:1999ji,Berti:2009kk,Konoplya:2011qq}.
Gravitational and electromagnetic perturbations appeared when studying 
Schwarzschild and charged black holes~\cite{Regge:1957td,Zerilli:1970se,Zerilli:1974ai,Moncrief:1974am}.
With the time dependence $e^{-\I\omega t}$, a negative imaginary
part of $\omega$ describes decay.
For asymptotically flat black holes, QNMs are usually defined
by outgoing waves at spatial infinity.
In AdS spacetime, the timelike conformal boundary
instead requires an additional boundary condition and can
reflect perturbations back toward the black hole.
The AdS boundary conditions thus play an important role in the
QNM spectrum, particularly for coupled perturbations~
\cite{Horowitz:1999jd,Cardoso:2001bb,Berti:2003ud,Wang:2004bv,
IshibashiWald2004,Liu:2025ipr}.

The Euler--Heisenberg (EH) interaction introduces
nonlinear corrections to Maxwell electrodynamics associated
with quantum electrodynamics~\cite{Heisenberg:1936nmg}.
Such corrections modify charged black-hole
solutions~\cite{Yajima:2000kw,Ruffini:2013hia} and affect
electromagnetic wave propagation and
stability~\cite{Obukhov:2002xa,Moreno:2002gg,Nomura:2020tpc,Nomura:2021efi}.
A famous charged black hole with  dilaton hair were found as the
Gibbons--Maeda and Garfinkle--Horowitz--Strominger
solution without EH interaction~\cite{Gibbons:1987ps,Garfinkle:1990qj}.
Bakopoulos \textit{et al.} obtained an exact magnetically charged
dilaton--Euler--Heisenberg (dEH) black hole and its AdS
extension~\cite{Bakopoulos:2024hah}, with a cosmological dilaton
potential~\cite{Gao:2004tu}.
The dEH model thus combines nontrivial dilaton
hair with nonlinear electromagnetic interactions.
The existence of an exact magnetic black-hole solution
provides a useful setting for examining how these interactions
affect both the background geometry and the coupled
perturbation spectrum.
In particular, the nonlinear electromagnetic contribution
can be distinguished from the dilatonic sector by comparison
with the solution at vanishing EH coupling.
Electromagnetic  and axial gravitational QNMs have been studied on
the magnetic dEH background~\cite{Zhang:2025xqt,Zou:2025rbu}.
Related studies have also examined EH black holes in de Sitter
spacetime and black holes surrounded by
matter~\cite{Zhang:2026nog,Zhang:2026yfb,Feng:2026yhj}.

However, perturbations of the dilaton hair generally couple
to metric and electromagnetic perturbations, as shown in
earlier studies of charged dilaton black
holes~\cite{Ferrari:2000ep,Brito:2018hjh}.
On a magnetic background, the physically indispensable sector is the Type-II sector, which contains polar gravitational and dilaton perturbations coupled to axial electromagnetic perturbations~
\cite{Nomura:2020tpc,Li:2026bek}.
Polar metric--dilaton perturbations and the full Type-II system have been studied for asymptotically flat dEH
black holes~\cite{Li:2026gqi,Li:2026bek}, while the
complementary Type-I sector contains axial gravitational
and polar electromagnetic perturbations~
\cite{Wang:2026iew}.

The AdS extension, however, involves more than
a change in the asymptotic geometry.
The cosmological dilaton potential contributes explicitly
to the background and linearized field equations and
determines the asymptotic effective mass of the dilaton.
In the present model, the relation
$m_{\mathrm{eff}}^2L^2=-2$ allows two distinct scalar falloffs
at the AdS boundary, making the choice of scalar boundary
condition part of the QNM problem~
\cite{Gao:2004tu,KlebanovWitten1999}.
A consistent analysis of the Type-II spectrum
therefore requires both the complete field couplings
and the appropriate boundary conditions for the
gravitational, dilaton, and electromagnetic perturbations.

Beyond dynamical stability, QNM frequencies serve as a powerful tool to probe thermodynamic phase transitions in AdS black holes. Quasinormal modes of minimally coupled test scalars have also
been used to probe the small- and large-black-hole
thermodynamic branches of charged AdS black holes,
including examples in massive gravity and
Gauss--Bonnet gravity~
\cite{Liu:2014gvf,Zou:2017juz,Zhang:2020khz}.
For dEH--AdS black holes, the relation between
a first-order phase transition and the QNMs of an external
massless scalar was recently studied
in~\cite{Chen:2026mxq}.
In this work, we instead consider the $\ell=0$
perturbation of the dynamical dilaton, whose reduced
equation incorporates the metric constraints rather
than treating the scalar as an external test field.
This allows us to examine the dilaton response
along the small- and large-black-hole branches of
a first-order phase transition under the chosen
AdS boundary conditions.

In the present work, we derive the Type-II perturbation
equations for $\ell=0,1,2$ modes of magnetically charged
dEH-AdS black holes, retaining the contributions from the
cosmological dilaton potential.
We examine how the nonlinear electromagnetic
interaction and AdS curvature affect the scalar-,
electromagnetic-, and gravitational-led QNM branches.
In particular, we compare the dipole frequencies with
those obtained at vanishing EH coupling to identify
the nonlinear corrections and their weak-charge behavior.
We also investigate the charge and curvature dependence
of the quadrupole modes, the approach of the dipole
frequencies to the small-AdS-black-hole
limit~\cite{Konoplya:2002zu}, and the dynamical
stability of the computed modes.
Finally, we study the dilaton monopole QNMs along
the two thermodynamic branches near the first-order
phase transition.
The QNM frequencies are calculated using direct
integration and Chebyshev pseudospectral
methods~\cite{Pani:2013pma,Jansen:2017oag,Lin:2016sch},
with ingoing boundary conditions at the horizon
and fixed independent perturbative sources at
the AdS boundary.

The remainder of this paper is organized as follows.
Section~\ref{sec:model} introduces the dEH-AdS black-hole background
and its allowed parameter region.
In Sec.~\ref{sec:perturbations}, we derive the Type-II perturbation equations,
while Sec.~\ref{sec:methods} presents the boundary conditions and
numerical methods.
The QNM results are presented in Sec.~\ref{sec:results},
followed by conclusions in Sec.~\ref{sec:conclusions}.
Additional details are provided in the appendices.

\section{dEH-AdS black hole and its allowed region}
\label{sec:model}

In geometrized units $G=c=1$, the action of dEH gravity with a
cosmological dilaton potential $V_\Lambda(\phi)$ is given by~\cite{Bakopoulos:2024hah,Gao:2004tu}
\begin{equation}
S=\frac{1}{16\pi}\int\mathrm{d}^4x\sqrt{-g}\Big[
R-2\nabla_\mu\phi\nabla^\mu\phi-e^{-2\phi}F^2
-f(\phi)\big(2\alpha F^\mu{}_{\nu}F^\nu{}_{\rho}F^\rho{}_{\delta}F^\delta{}_{\mu}
-\beta F^4\big)-V_\Lambda(\phi)\Big],
\label{eq:action}
\end{equation}
\noindent where $R$ is the Ricci scalar,
$F_{\mu\nu}=\partial_\mu A_\nu-\partial_\nu A_\mu$ is the
electromagnetic field strength, $F^2=F_{\mu\nu}F^{\mu\nu}$ and
$F^4=(F^2)^2$.
The parameters $\alpha$ and $\beta$ denote the nonlinear electromagnetic (NED)
couplings, and the dilaton coupling function and dilaton potential take the forms
\begin{equation}
f(\phi)=-3\cosh(2\phi)-2,\qquad
V_\Lambda(\phi)=\frac{2\Lambda}{3}\big[\cosh(2\phi)+2\big].
\label{eq:potential}
\end{equation}
For $\Lambda<0$, the magnetic charged black-hole
solution can be expressed  as~\cite{Bakopoulos:2024hah}
\begin{align}
\mathrm{d}s^2&=-A(r)\mathrm{d}t^2+\frac{\mathrm{d}r^2}{B(r)}
+r^2\mathrm{d}\Omega_2^2,\label{eq:metric}\\
A(r)&=1-\frac{4}{Q^2+\sqrt{Q^4+4r^2}}
-\frac{2\zeta Q^4}{r^6}-\frac{\Lambda r^2}{3},\nonumber\\
B(r)&=A(r)\frac{Q^4+4r^2}{4r^2},\qquad
\bar\phi(r)=\operatorname{arcsinh}\!\left(\frac{Q^2}{2r}\right),
\label{eq:background}
\end{align}
\noindent where $Q=Q_m/M$, $\zeta=(\alpha-\beta)/M^2$ and
$\Lambda=M^2\Lambda_{\rm phys}$, with $Q_m$ and $M$ denoting the magnetic
charge and black-hole mass, respectively.
The coordinates are measured in units of $M$, and $M=1$ is set
in the following calculation, with frequencies denoted by $M\omega$.
The background magnetic potential and field strength are chosen to be
\begin{equation}
A_\mu\mathrm{d}x^\mu=Q\cos\theta\,\mathrm{d}\varphi,\qquad
F_{\theta\varphi}=-Q\sin\theta,
\label{eq:magnetic-field}
\end{equation}
\noindent where $F_{\theta\varphi}=\partial_\theta A_\varphi$, and the sign
convention follows Ref.~\cite{Li:2026bek}.
$V_\Lambda(0)=2\Lambda$ implies that the asymptotic  geometry is AdS
with negative $\Lambda$.
We note that the solution recovers the Schwarzschild--AdS black hole when $Q=0$
and asymptotically flat dEH black hole for $\Lambda=0$.
The event horizon $r_h(Q,\zeta,\Lambda)$ is determined by the largest positive root of $A(r_h)=0$.

\begin{figure}[tbp]
\centering
\includegraphics[width=0.55\linewidth]{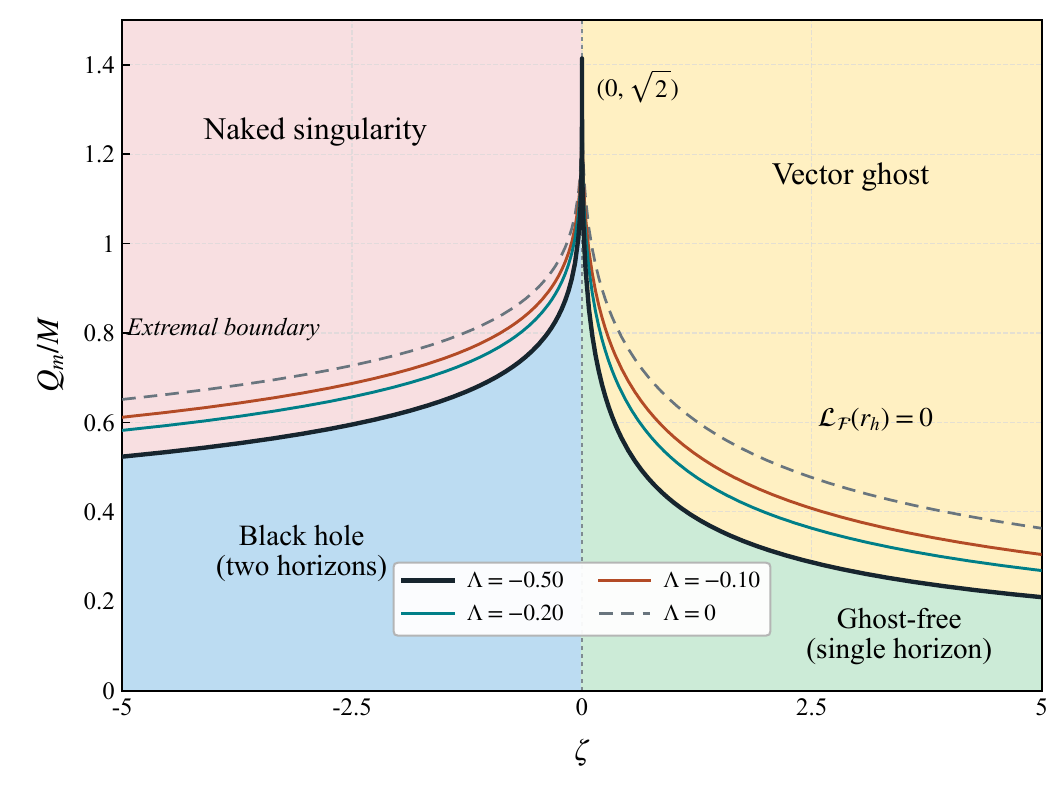}
\caption{Ghost free-region in parameter space for 
$-5\leq\zeta=(\alpha-\beta)/M^2\leq5$ vs  $Q_m/M$ with $M=1$.
The curves correspond to $\Lambda=0,-0.10,-0.20,-0.50$, as identified
by the legend inside the plot. The shaded boundary refers to $\Lambda=-0.50$.
The negative coupling boundary denotes  extremal limit, whereas the positive coupling
boundary is determined by  ${\cal L}_{\cal F}(r_h)=0$.}
\label{fig:ads-ghost-phase}
\end{figure}

Let us explore the kinetically allowed region for electromagnetic propagation because the action includes NED terms. 
For the purely magnetic background, one has 
$F^\mu{}_{\nu}F^\nu{}_{\rho}F^\rho{}_{\delta}F^\delta{}_{\mu}
=\mathcal F^2/2$, with $\mathcal F=F_{\mu\nu}F^{\mu\nu}$.
The electromagnetic part of the Lagrangian in \eqref{eq:action}
thus reduces to
$\mathcal L_{\rm em}=-e^{-2\phi}\mathcal F-\zeta f(\phi)\mathcal F^2$.
With this sign convention, the necessary condition for the absence
of an electromagnetic ghost is given by~\cite{Li:2026bek,Li:2026gqi}
\begin{equation}
\begin{gathered}
{\cal L}_{\cal F}\equiv
-\left.\frac{\partial\mathcal L_{\rm em}}{\partial\mathcal F}\right|_{\rm bg}
=e^{-2\bar\phi}+2\zeta f(\bar\phi)\bar{\mathcal F}>0,
\qquad \bar{\mathcal F}=\frac{2Q^2}{r^4}.
\end{gathered}
\label{eq:ghost}
\end{equation}
The condition \eqref{eq:ghost} is displayed in parameter space (see Fig.~\ref{fig:ads-ghost-phase}).
Although the dilaton potential does not contribute directly to
the electromagnetic kinetic term, it affects  the horizon radius and
therefore, the exterior region on which this condition is imposed.
For $\zeta<0$, the limiting charge is determined by the merger of two
horizons, beyond which a regular black-hole exterior no longer exists.
For $\zeta>0$, the condition of ${\cal L}_{\cal F}(r_h)=0$ separates the
ghost-free single-horizon solutions from the vector-ghost region.
The $\Lambda=0$ curve implies  the corresponding asymptotically flat
boundary for comparison, showing the largest ghost-free region.

Finally, we note that the kinetic condition eliminates electromagnetic ghosts. But the
dynamical stability of dEH-AdS black holes is determined by solving  their
coupled evolution equations on the ghost-free region.

\FloatBarrier

\section{Coupled perturbation equations}
\label{sec:perturbations}

In the Type-II sector, polar gravitational and dilaton perturbations
are coupled to axial electromagnetic perturbations on the dEH-AdS
background~\cite{Nomura:2020tpc,Li:2026bek}.
The fields are expanded to first order perturbations  as
\begin{equation}
g_{\mu\nu}=\bar g_{\mu\nu}+\epsilon h_{\mu\nu},\qquad
\phi=\bar\phi+\epsilon\delta\phi,\qquad
A_\mu=\bar A_\mu+\epsilon\delta A_\mu,\label{eq:perturbations}
\end{equation}
\noindent where overbars denote the static background fields and $\epsilon$ is regarded as a small expansion parameter.
Taking the time dependence $e^{-\I\omega t}$ and adopting the
Regge--Wheeler gauge~\cite{Regge:1957td}, we can write the perturbations as
\begin{align}
\delta\phi&=e^{-\I\omega t}\frac{\phi_1(r)}rY_\ell^m,
\label{eq:scalar-pert}\\
h_{\mu\nu}&=e^{-\I\omega t}Y_\ell^m
\begin{pmatrix}
AH_0&H_1&0&0\\
H_1&H_2/B&0&0\\
0&0&r^2K_p&0\\
0&0&0&r^2\sin^2\theta K_p
\end{pmatrix},\label{eq:metric-pert}\\
\delta A_\mu&=e^{-\I\omega t}
\left(0,0,-\frac{u_4\partial_\varphi Y_\ell^m}{\sin\theta},
u_4\sin\theta\partial_\theta Y_\ell^m\right).
\label{eq:em-pert}
\end{align}
\noindent Here $Y_\ell^m$ are scalar spherical harmonics and $K_p$ is the polar
angular metric amplitude, with frequency integrals and harmonic sums
omitted for simplicity.
We note that spherical symmetry makes the radial equations independent of $m$.
The dilaton potential contributes to both the linearized
Einstein and scalar equations, and  varying the corresponding terms
in the bilinearized action gives two linearized Einstein and dilaton equations as 
\begin{align}
\delta{\cal E}_{\mu\nu}\big|_V
&=\frac12\bar V h_{\mu\nu}
+\frac12\bar g_{\mu\nu}V_1\delta\phi,\nonumber\\
\delta{\cal S}\big|_V&=-V_2\delta\phi,
\label{eq:potential-variation}
\end{align}
\noindent where the dilaton  equation is normalized with leading term $4\Box\phi$
and its potentials are given by
\begin{equation}
\bar V=V_\Lambda(\bar\phi),\qquad
V_1=\frac{4\Lambda}{3}\sinh(2\bar\phi),\qquad
V_2=\frac{8\Lambda}{3}\cosh(2\bar\phi).
\label{eq:potential-derivatives}
\end{equation}
\noindent The generalized Maxwell equation contains no explicit variation of
$V_\Lambda$, while the terms in \eqref{eq:potential-variation} must
be retained in the gravitational and scalar linearized equations.

\subsection{\texorpdfstring{$\ell=0$}{ell=0} mode: dilaton}
\label{subsec:l0}

For $\ell=0$, the axial vector and trace-free tensor harmonics vanish,
and the metric perturbations are determined by the constraints at fixed
magnetic charge $Q$.
The dilaton is therefore the only physical  degree of freedom.
Eliminating the metric amplitudes gives the radial equation
\begin{equation}
\phi_1''+P\phi_1'+\left(\frac{\omega^2}{AB}+{\cal U}_0\right)\phi_1=0,
\qquad P=\frac12\left(\frac{A'}A+\frac{B'}B\right),
\label{eq:l0}
\end{equation}
\noindent where primes denote derivatives with respect to $r$.
The coefficient ${\cal U}_0$ is given by
\begin{align}
{\cal U}_0={}&-\frac{\Lambda}{3B}(e^{-2\bar\phi}+e^{2\bar\phi})
-\frac{2Q^2e^{-2\bar\phi}}{r^4B}
+\frac{6\zeta Q^4}{r^8B}(e^{-2\bar\phi}+e^{2\bar\phi})
-\frac{A'}{2rA}-\frac{B'}{2rB}\nonumber\\
&+\frac{\Lambda r\bar\phi'}{6B}(e^{-2\bar\phi}-e^{2\bar\phi})
+\frac{Q^2e^{-2\bar\phi}\bar\phi'}{r^3B}
+\frac{3\zeta Q^4\bar\phi'}{r^7B}(e^{2\bar\phi}-e^{-2\bar\phi})\nonumber\\
&+2(\bar\phi')^2-\frac{3rA'}{2A}(\bar\phi')^2
+\frac{rB'}{2B}(\bar\phi')^2-r^2(\bar\phi')^4.
\label{eq:l0-coefficient}
\end{align}
Introducing the tortoise coordinate through
$\mathrm{d}r_*/\mathrm{d}r=1/\sqrt{AB}$ and multiplying
\eqref{eq:l0} by $AB$ gives the monopole dilaton equation in the
Schr\"odinger-like form
\begin{equation}
\frac{\mathrm{d}^2\phi_1}{\mathrm{d}r_*^2}
+\big[\omega^2-V_0(r)\big]\phi_1=0,\qquad V_0=-AB{\cal U}_0,
\label{eq:l0-master}
\end{equation}
\noindent where the effective potential $V_0$ contains the cosmological
contribution through the background functions and through the first
and second derivatives of $V_\Lambda$.

\subsection{\texorpdfstring{$\ell=1$}{ell=1} mode: dilaton--electromagnetic sector}
\label{subsec:l1}

For $\ell=1$, the trace-free polar tensor harmonic vanishes.
With $K_p=0$ and $H_2=H_0$, the remaining metric amplitudes can be
eliminated by using the Einstein constraints, leaving the dilaton
amplitude $\phi_1$ and electromagnetic amplitude $u_4$ as the two
propagating fields.
Following Ref.~\cite{Li:2026bek}, the coefficient functions are defined as
\begin{align}
{\cal C}=3+4e^{2\bar\phi}+3e^{4\bar\phi},\quad
{\cal D}=r^4-2\zeta Q^2{\cal C},\quad
{\cal E}=r^4+6\zeta Q^2(e^{4\bar\phi}-1).
\label{eq:shortdefs}
\end{align}
Two coupled scalar and electromagnetic equations take the forms
\begin{align}
0={}&\phi_1''+P\phi_1'+{\cal V}_1\phi_1
+\frac{4Qe^{-2\bar\phi}{\cal E}}{r^7B}u_4
+\frac{4Qe^{-2\bar\phi}\bar\phi'{\cal D}}{r^5}u_4',\label{eq:l1-scalar}\\
0={}&u_4''+{\cal P}_u u_4'
+\left[\frac{\omega^2}{AB}+\frac{-2r^4+12\zeta Q^2{\cal C}}{Br^2{\cal D}}\right]u_4
+\frac{2Q{\cal E}}{r^3B{\cal D}}\phi_1,
\label{eq:l1-em}
\end{align}
where 
\begin{align}
{\cal V}_1={}&\frac{\omega^2}{AB}-\frac4{r^2}
+\frac{4A}{r^2B}-\frac{A'}{2rA}-\frac{B'}{2rB}
-\frac2{r^2B}-\frac{2Q^2e^{-2\bar\phi}}{r^4B}
+\frac{6\zeta Q^4}{r^8B}(e^{-2\bar\phi}+e^{2\bar\phi})
-\frac{\Lambda}{3B}(e^{-2\bar\phi}+e^{2\bar\phi}),\nonumber\\
{\cal P}_u={}&\frac12\left[\frac{A'}A+\frac{B'}B
+\frac{-4r^5\bar\phi'-8\zeta Q^2\{3r(e^{4\bar\phi}-1)\bar\phi'-2{\cal C}\}}{r{\cal D}}\right].
\nonumber
\end{align}
It is clear that scalar and electromagnetic equations are  decoupled for $Q=0$.
At nonzero charge, the terms proportional to $Q$ couple $\phi_1$
and $u_4$, so each dipole mode contains both fields.
The cosmological dilaton potential  appears as the term $-V_2/(4B)$ in
${\cal V}_1$, in addition to its effect on the background geometry.

\subsection{\texorpdfstring{$\ell=2$}{ell=2} mode: gravitational--dilaton--electromagnetic sector}
\label{subsec:l2}

For $\ell=2$, polar gravitational perturbation is present together with
the dilaton and axial electromagnetic perturbations.
As in Ref.~\cite{Li:2026bek}, taking $H_2=H_0$ and
$H_1=\omega R_h$ leaves the metric amplitude vector $(K_p,R_h,H_0)$.
The coupled system consists of three first-order Einstein equations,
one algebraic Einstein constraint and two second-order matter equations.
For convenience, we define
\begin{equation}
{\cal G}=\frac{A-2B-rB'-2}{r^2}
-\frac{Q^2e^{-2\bar\phi}}{r^4}+\frac{\zeta Q^4e^{-2\bar\phi}{\cal C}}{r^8}
-\frac{\bar V}{2},\qquad
{\cal H}=r^4-6\zeta Q^2{\cal C}.
\label{eq:l2-background-combinations}
\end{equation}
\noindent The first-order gravitational equations are given by
\begin{align}
0={}&K_p'+\left(\frac1r-\frac{A'}{2A}\right)K_p
-\frac{H_0}{r}+\frac{2\bar\phi'}{r}\phi_1+\I{\cal G}R_h,
\label{eq:l2-eqK}\\
0={}&R_h'+PR_h+\frac{\I}{B}(H_0+K_p)
-\frac{4\I Qe^{-2\bar\phi}{\cal D}}{r^6B}u_4,
\label{eq:l2-eqR}\\
0={}&H_0'-\left(\frac1r-\frac{A'}A\right)H_0
-\left(\frac{A'}{2A}-\frac1r\right)K_p
+\I\left({\cal G}+\frac{\omega^2}A\right)R_h\nonumber\\
&+\frac{4Qe^{-2\bar\phi}{\cal D}}{r^6}u_4'-\frac{2\bar\phi'}r\phi_1.
\label{eq:l2-eqH0}
\end{align}
The term $-\bar V/2$ in ${\cal G}$ comes from the explicit variation
of the cosmological potential in the Einstein equations.
Eliminating the metric derivatives from the $\{rr\}$ component gives the
algebraic constraint
\begin{equation}
0=C_HH_0+C_RR_h+C_KK_p+C_u u_4+C_{u'}u_4'
+C_\phi\phi_1+C_{\phi'}\phi_1',
\label{eq:l2-constraint}
\end{equation}
\noindent where the coefficient functions are given in appendix~\ref{app:operator}.
When $C_H\ne0$, this constraint determines $H_0$ in terms of the
remaining amplitudes. We note that  this condition holds throughout the open exterior
at the reference quadrupole parameters, as shown in the appendix.
The evolution equations for $K_p$ and $R_h$ are then solved with
$H_0$ reconstructed from the constraint, while Eq.\eqref{eq:l2-eqH0}
provides an independent check.
The axial electromagnetic equation takes the form
\begin{align}
0={}&u_4''+{\cal P}_u u_4'
+\left(\frac{\omega^2}{AB}
-\frac{6{\cal H}}{Br^2{\cal D}}\right)u_4
+\frac{Q{\cal H}}{r^2B{\cal D}}K_p
+\frac{2Q{\cal E}}{r^3B{\cal D}}\phi_1,
\label{eq:l2-em}
\end{align}
\noindent where ${\cal P}_u$ is the derivative coefficient in \eqref{eq:l1-em}.
Substituting the metric evolution equations into the dilaton equation
leads to 
\begin{align}
0={}&\phi_1''+P\phi_1'+{\cal V}_2\phi_1
-\frac{2Q^2e^{-2\bar\phi}{\cal E}}{r^7B}K_p
+\frac{12Qe^{-2\bar\phi}{\cal E}}{r^7B}u_4
+\frac{4Qe^{-2\bar\phi}\bar\phi'{\cal D}}{r^5}u_4',
\label{eq:l2-scalar}\\
{\cal V}_2={}&\frac{\omega^2}{AB}
-\frac{A'}{2rA}-\frac{B'}{2rB}
-\frac6{r^2B}-4(\bar\phi')^2-\frac{2Q^2e^{-2\bar\phi}}{r^4B}\nonumber\\
&+\frac{6\zeta Q^4}{r^8B}(e^{-2\bar\phi}+e^{2\bar\phi})-\frac{V_2}{4B}.
\label{eq:l2-scalar-coefficient}
\end{align}
\noindent The algebraic constraint reduces the number of independent local
integration constants from seven to six, so the physical state can
be composed of  
\begin{equation}
\bm\Psi_2=(K_p,R_h,\phi_1,\phi_1',u_4,u_4')^{\mathsf T}.
\label{eq:l2-physical-state}
\end{equation}
\noindent This system describes three coupled physical degrees of freedom.
That is, gravitational-, scalar- and electromagnetic-led modes are
identified by their field content in the weakly coupled regime and
followed continuously as the charge increases.

\section{AdS boundary conditions and numerical methods}
\label{sec:methods}
In this section, we briefly describe boundary conditions and numerical methods to compute QNM frequencies of each mode. 
The radial equations will be solved by  imposing on ingoing
condition at the event horizon and vanishing perturbative sources
at the AdS boundary. 
The direct integration and Chebyshev pseudospectral calculations
will be used under the same boundary conditions for each multipole mode.

\subsection{AdS boundary conditions}
\label{subsec:boundary}

For a nonextremal horizon, the surface gravity is
\begin{equation}
\kappa=\frac12\sqrt{A'(r_h)B'(r_h)}.
\label{eq:surface-gravity}
\end{equation}
\noindent The tortoise coordinate then behaves as
$r_*\simeq\ln(r-r_h)/(2\kappa)$ near the future horizon,
so the ingoing radial factor takes the form
\begin{equation}
e^{-\I\omega r_*}\simeq(r-r_h)^{\lambda_h},\qquad
\lambda_h=-\frac{\I\omega}{2\kappa}.
\label{eq:ingoing}
\end{equation}
\noindent The derivative and metric components contain additional powers of
$r-r_h$, which are removed before constructing the regular endpoint
expansions.
The ingoing solutions then have one, two and three independent
amplitudes for $\ell=0,1,2$, respectively.

The AdS boundary is reached at finite $r_*$ as $r\to\infty$, and
its timelike character requires a boundary condition for well-defined
evolution~\cite{IshibashiWald2004}.
For the dilaton, the asymptotic linear equation gives an effective
mass $m_{\rm eff}^2=V_\Lambda''(0)/4=2\Lambda/3$, so the two
AdS$_4$ exponents satisfy
\begin{equation}
\Delta(\Delta-3)=-\frac{3m_{\rm eff}^2}{\Lambda}=-2,
\qquad \Delta=1,2.
\label{eq:scalar-delta}
\end{equation}
Choosing $\delta\phi=\phi_1/r$, its expansion reads as 
\begin{equation}
\phi_1=a_0+\frac{a_1}r+\cdots,\qquad
\delta\phi=\frac{a_0}r+\frac{a_1}{r^2}+\cdots.
\label{eq:scalar-infinity}
\end{equation}
\noindent Here, with $L=\sqrt{-3/\Lambda}$, the effective mass satisfies
\begin{equation}
    m_{\rm eff}^2L^2=-2>-\frac{9}{4}=m_{\rm BF}^2L^2.
\end{equation}
Thus, the AdS$_4$ Breitenlohner--Freedman bound~\cite{Breitenlohner:1982jf}
is satisfied throughout the parameter space considered,
independently of $Q$ and $\zeta$.
We adopt the standard scalar quantization with $\Delta=2$ and impose
$a_0=0$, which selects the faster falloff of the dilaton
perturbation~\cite{KlebanovWitten1999}.
The background behaves as $\bar\phi=Q^2/(2r)+O(r^{-3})$,
whose slow coefficient changes with $Q$ but it is independent of
$\zeta$ at fixed mass and charge.

For the electromagnetic amplitude, the asymptotic expansion shows
$u_4=b_0+b_1/r+\cdots$, and fixing the boundary gauge potential
requires $b_0=0$.
For a quadrupole metric amplitude, one has $K_p=k_0+O(r^{-1})$ and the
fixed boundary metric indicates  $k_0=0$.
The independent source conditions are therefore $k_0=a_0=b_0=0$,
while the other metric amplitudes  are determined by the Einstein
equations and constraint.
The corresponding source matrix and asymptotic solutions in the
physical fields are given in
Appendix~\ref{app:boundary-subspaces}~\cite{Liu:2025ipr}.

\subsection{Direct integration (DI) method}
\label{subsec:di}

For the monopole case, we integrate \eqref{eq:l0} for $\phi_1$,
while the dipole requires two coupled equations
\eqref{eq:l1-scalar} and \eqref{eq:l1-em} for $(\phi_1,u_4)$.
For the quadrupole, we first eliminate $H_0$ with
\eqref{eq:l2-constraint} and integrate \eqref{eq:l2-eqK},
\eqref{eq:l2-eqR}, \eqref{eq:l2-em} and \eqref{eq:l2-scalar}.
Introducing the radial derivatives of the matter fields, these
equations can be written as
\begin{equation}
\bm\Psi_\ell'=\mathsf A_\ell(r,\omega)\bm\Psi_\ell,\qquad
\bm\Psi_0=(\phi_1,\phi_1')^{\mathsf T},\qquad
\bm\Psi_1=(\phi_1,\phi_1',u_4,u_4')^{\mathsf T},
\label{eq:di-first-order}
\end{equation}
\noindent where $\bm\Psi_2$ is given in
\eqref{eq:l2-physical-state}, and $\mathsf A_\ell$ is obtained
directly from solving  the corresponding radial equations, with the background
quantities in \eqref{eq:background}.
Thus, the integrated systems have two, four and six components for
$\ell=0,1,2$, respectively.

The endpoint solutions are constructed by substituting Frobenius
series into the radial equations~\cite{Leaver:1985ax,Pani:2013pma}.
For the dipole, the ingoing matter amplitudes take the form
\begin{equation}
\begin{pmatrix}\phi_1\\u_4\end{pmatrix}
=(r-r_h)^{\lambda_h}\sum_{n=0}^{N_h}
\begin{pmatrix}a_n^{(h)}\\b_n^{(h)}\end{pmatrix}(r-r_h)^n,
\label{eq:di-horizon-series}
\end{equation}
\noindent where $\lambda_h$ is defined in \eqref{eq:ingoing}.
The two choices $(a_0^{(h)},b_0^{(h)})=(1,0)$ and $(0,1)$ generate the
independent ingoing solutions.  The coupled equations determine all
higher coefficients.
The monopole case has only such a solution, while the quadrupole has
three after the Einstein constraint is imposed.
At infinity, we retain the $r^{-1}$ matter solutions and, for the
quadrupole, the three source-free solutions derived in
Appendix~\ref{app:boundary-subspaces}.
All derivative and metric components are obtained from the radial
equations rather than assigned separate boundary data.

The horizon solutions are integrated outwards from
$r_h+\epsilon_h$, and the boundary solutions inwards from a large
radius $r_i$, using an adaptive Runge--Kutta method.
At a common radius $r_m$, they form the $2d\times d$ matrices
$S_h$ and $S_i$, where $d=1,2,3$ for the three multipoles.
A QNM is allowed to  exist when a linear combination of the ingoing solutions
matches a source-free solution in every state component, giving
\begin{equation}
Y(\omega)=[S_h(r_m),-S_i(r_m)],\qquad
F(\omega)=\det(D_{\rm row}YD_{\rm col})=0.
\label{eq:matching-matrix}
\end{equation}
\noindent Here, $D_{\rm row}$ and $D_{\rm col}$ are nonsingular
diagonal matrices used to normalize the rows and columns.
For $\ell=0$, this condition reduces to matching $\phi_1$ and
$\phi_1'$, while for $\ell=1,2$, it matches all four or six components.
The real and imaginary parts of the matching condition are solved
with a nearby frequency of the same mode as the initial guess.
The QNM frequency is checked by varying $N_h$, the boundary expansion
order, $\epsilon_h$, $r_i$ and $r_m$.
For the quadrupole, we also reconstruct $H_0$ and substitute it
into \eqref{eq:l2-eqH0} to check the eliminated Einstein equation.

\subsection{Pseudospectral (PS) method}
\label{subsec:ps}

We are aware that the monopole and dipole equations are also solved by Chebyshev
collocation~\cite{Canuto2006,Fornberg1996,Trefethen2000}.
To remove the square root from the background functions, we
introduce a new variable $y$ as 
\begin{equation}
y=\frac{Q^2+\sqrt{Q^4+4r^2}}{2},\qquad
r=\sqrt{y(y-Q^2)},\qquad y>Q^2.
\label{eq:y-map}
\end{equation}
The background expressed in terms of this coordinate is given by 
\begin{align}
A_y&=1-\frac2y-\frac{2\zeta Q^4}{y^3(y-Q^2)^3}
-\frac\Lambda3y(y-Q^2),\nonumber\\
B_y&=A_y\frac{(2y-Q^2)^2}{4y(y-Q^2)},\qquad
\bar\phi_y=\frac12\ln\frac{y}{y-Q^2}.\label{eq:y-background}
\end{align}
\noindent Here, $A_y=A(r(y))$, $B_y=B(r(y))$, and the radial
metric component becomes $(\mathrm dr/\mathrm dy)^2/B_y=1/A_y$.
A further transformation to a compact coordinate $x$ shows
\begin{equation}
x=1-\frac{y_h}{y},\qquad 0\leq x\leq1,\qquad y_h=y(r_h),
\label{eq:ps-compact-map}
\end{equation}
\noindent which  maps the horizon to $x=0$ and the AdS boundary to $x=1$.
In the radial equations, we have relations 
$\partial_r=\xi\partial_x$ and
$\partial_r^2=\xi^2\partial_x^2+\xi\xi_x\partial_x$
with $\xi=(y_h/y^2)\mathrm dy/\mathrm dr$.

For the dipole case, both $\phi_1$ and $u_4$ have the same ingoing
factor and the source-free $1/r$ falloff, so we  may write
\begin{equation}
\begin{pmatrix}\phi_1\\u_4\end{pmatrix}
=x^{-\I\omega/(2\kappa)}(1-x)
\begin{pmatrix}\widehat\phi_1\\\widehat u_4\end{pmatrix}.
\label{eq:ps-factor}
\end{equation}
The regular amplitudes are represented on the open
Chebyshev--Gauss grid
$x_j=[1+\cos((2j-1)\pi/(2N))]/2$, $j=1,\ldots,N$.
Substituting \eqref{eq:ps-factor} into \eqref{eq:l1-scalar}
and \eqref{eq:l1-em}, and replacing the derivatives of the
regular amplitudes by the Chebyshev differentiation matrices,
we obtain a compact matrix equation as 
\begin{equation}
\begin{gathered}
(\mathsf A_0+\omega\mathsf A_1+\omega^2\mathsf A_2)\bm a=0,
\\
\bm a=(\widehat\phi_1(x_1),\ldots,\widehat\phi_1(x_N),
\widehat u_4(x_1),\ldots,\widehat u_4(x_N))^{\mathsf T}.
\end{gathered}
\label{eq:pep}
\end{equation}
\noindent The off-diagonal blocks contain the scalar--electromagnetic
couplings in the two radial equations, including  $u_4'$ term
in \eqref{eq:l1-scalar}.
The frequency dependence remains quadratic after removing the ingoing factor and the source conditions are already imposed by the
factor $1-x$.
The companion generalized eigenvalue problem can be  solved, and each
retained eigen-pair is substituted back into \eqref{eq:pep} to check
its residual~\cite{TisseurMeerbergen2001,Jansen:2017oag,Lin:2016sch}.

For the monopole,  substituting 
$\phi_1=e^{-\I\omega r_*}\psi$ into \eqref{eq:l0-master} gives
\begin{equation}
AB\psi''+\left[\frac12(AB)'-2\I\omega\sqrt{AB}\right]\psi'
-V_0\psi=0,
\label{eq:l0-ef-equation}
\end{equation}
\noindent which is linear in frequency $\omega$.
We discretize $\psi$ on a Chebyshev--Lobatto grid including
$x=0,1$.
The horizon row is replaced by the regular limit of this equation,
$4\kappa(\kappa-\I\omega)\psi_x(0)-v_h\psi(0)=0$,
where $v_h=\lim_{x\to0}V_0/x$ and the boundary row imposes
$\psi(1)=0$.
Then, the resulting matrix equation becomes a linear eigenvalue equation.

For the quadrupole case, we instead apply Chebyshev collocation to the
six-component system after eliminating $H_0$.
Writing $z=r-r_h$ and $\sigma=\ln z$, we may rescale the state as
\begin{equation}
\bm U=e^{-\lambda_h(\sigma-\sigma_0)}r\mathsf S\bm\Psi_2,
\qquad \mathsf S=\operatorname{diag}(1,z,1,z,1,z).
\label{eq:l2-ps-state}
\end{equation}
\noindent Here, $\sigma_0$ is the starting point of each integration
side and  the factors remove the known horizon powers and account for
the different falloffs of ($K_p$, $R_h$, $\phi_1$, $u_4$).
On each logarithmic radial interval, the transformed equations
are imposed at Chebyshev--Lobatto nodes, with the three Frobenius
solutions prescribed at the initial node and all six components
being continuous between adjacent intervals.
Solving these linear systems propagates the horizon and boundary
bases to $r_m$.
Restoring $\bm\Psi_2$ provides the matching condition
\eqref{eq:matching-matrix}.
The low-charge dipole comparisons also use collocation matching,
which is applied to the four-component dipole system in $x$.
For all multipoles, each mode is followed by its frequency and
normalized radial profile, with convergence checked by increasing
the spectral resolution and varying the endpoint expansions.

\FloatBarrier

\section{QNM frequencies}
\label{sec:results}

\label{subsec:coverage}

In this section, we investigate the charge and curvature
dependence of the Type-II QNMs.
The complex frequency is written as $\omega=\omega_R-\I\gamma$,
where $\omega_R$ is the oscillation frequency, $\gamma>0$ is
the damping rate, and $\tau=1/\gamma$ is the decay time.
We set $M=1$ unless otherwise specified.

\subsection{Monopole modes and representative frequencies}
\label{subsec:frequencies}

The dilaton  monopole frequencies at $\Lambda=-0.05$ are listed in
Table~\ref{tab:monopole-charge}.
As the charge increases, the dilaton perturbation oscillates
more rapidly and decays more slowly, as shown in
Table~\ref{tab:monopole-charge}.
The two signs of $\zeta$ give the almost same frequencies at small
charge, but a more visible difference at $Q=0.30$, as expected
from the $Q^4$ dependence of the NED terms in
Eq.\eqref{eq:l0-coefficient}.

\begin{table}[H]
\caption{Monopole dilaton  spectral frequencies based on PS and their absolute
differences from independent DI at $M=1$ and
$\Lambda=-0.05$.}
\label{tab:monopole-charge}
\centering
\begin{ruledtabular}
\begin{tabular}{cccc}
$\zeta$ & $Q$ & $M\omega_{\rm PS}$ & $|M\omega_{\rm PS}-M\omega_{\rm DI}|$\\
\hline
1 & 0.01 & $0.217139457402-0.032068628270\,\I$ & $2.71\times10^{-12}$\\
1 & 0.10 & $0.217209291984-0.031856834028\,\I$ & $2.67\times10^{-12}$\\
1 & 0.30 & $0.217796577018-0.030160826197\,\I$ & $2.66\times10^{-12}$\\
-1 & 0.01 & $0.217139457374-0.032068628216\,\I$ & $2.67\times10^{-12}$\\
-1 & 0.10 & $0.217209005307-0.031856285589\,\I$ & $2.67\times10^{-12}$\\
-1 & 0.30 & $0.217771411066-0.030111399758\,\I$ & $2.68\times10^{-12}$\\
\end{tabular}
\end{ruledtabular}
\end{table}

Table~\ref{tab:frequencies} gives six representative frequencies
for $\ell=0,1,2$ modes at $(\zeta,Q,\Lambda)=(1,0.01,-0.05)$.
Here GR-, SC- and EM- denote the gravitational-, scalar- and
electromagnetic-led modes, respectively, according to their field
content in the weakly coupled regime.
As the charge increases, the dilaton perturbation oscillates more rapidly and decays more slowly, as shown in Table~\ref{tab:monopole-charge}.

\begin{table}[tbp]
\caption{DI frequencies for $\ell=0,1,2$ at $M=1$, $\zeta=1$,
$Q=0.01$ and $\Lambda=-0.05$.}
\label{tab:frequencies}
\centering
\begin{ruledtabular}
\begin{tabular}{clc}
$\ell$ & -led mode & $M\omega_{\rm DI}$\\
\hline
0 & SC & $0.217139457403-0.032068628270\,\I$\\
1 & SC & $0.346542172502-0.008187736301\,\I$\\
1 & EM & $0.328654485716-0.015853614960\,\I$\\
2 & GR & $0.353865704330-0.000472469999\,\I$\\
2 & SC & $0.486639198654-0.000957608853\,\I$\\
2 & EM & $0.479078008834-0.001661229697\,\I$\\
\end{tabular}
\end{ruledtabular}
\end{table}

\FloatBarrier

\subsection{Coupled dipole modes}
\label{subsec:charge}

\begin{figure}[tbp]
\centering
\subfigure[\enspace SC-led mode]{\includegraphics[width=0.45\linewidth]{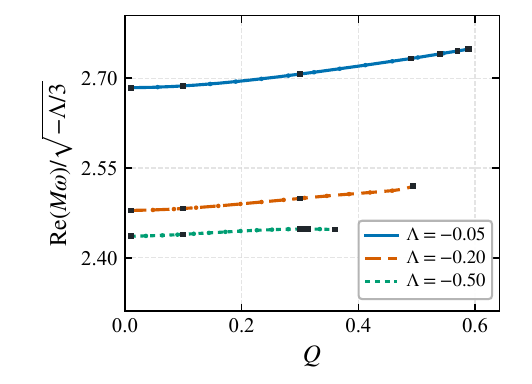}\label{fig:l1-plus-a}}
\hfill
\subfigure[\enspace SC-led mode]{\includegraphics[width=0.45\linewidth]{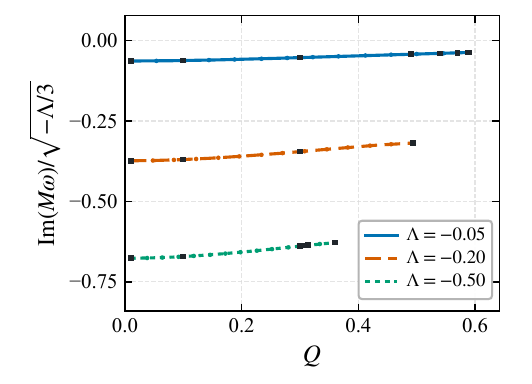}\label{fig:l1-plus-b}}
\\[2pt]
\subfigure[\enspace EM-led mode]{\includegraphics[width=0.45\linewidth]{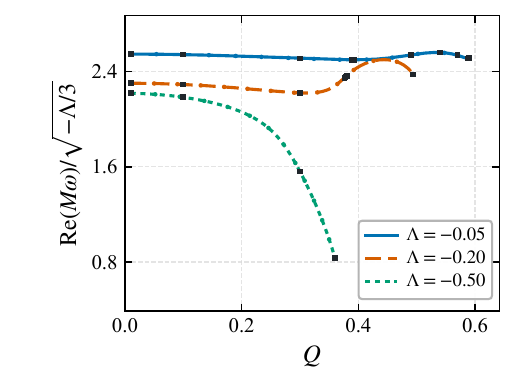}\label{fig:l1-plus-c}}
\hfill
\subfigure[\enspace EM-led mode]{\includegraphics[width=0.45\linewidth]{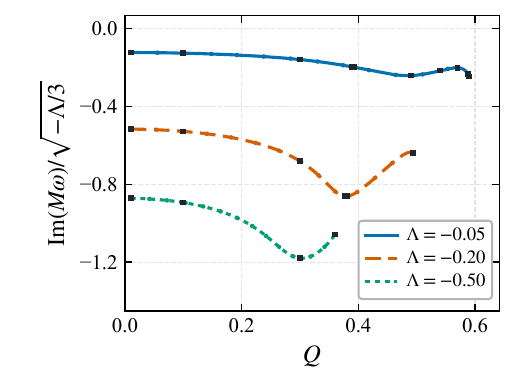}\label{fig:l1-plus-d}}
\caption{Scaled dipole frequencies with fixed $M=1$, $\zeta=1$ and
$\Lambda=-0.05,-0.20,-0.50$. The left and right columns show
$\operatorname{Re}(M\omega)/\sqrt{-\Lambda/3}$ and
$\operatorname{Im}(M\omega)/\sqrt{-\Lambda/3}$, respectively.}
\label{fig:l1-plus}
\end{figure}

\begin{figure}[tbp]
\centering
\subfigure[\enspace SC-led mode]{\includegraphics[width=0.40\linewidth]{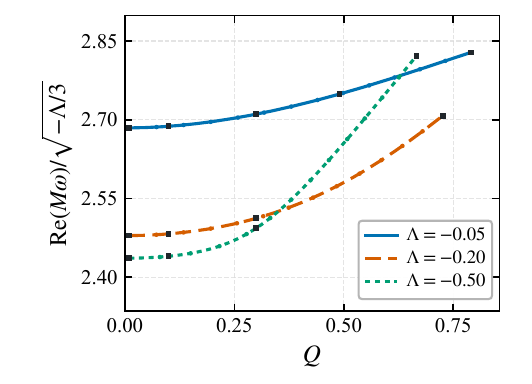}\label{fig:l1-minus-a}}
\hfill
\subfigure[\enspace SC-led mode]{\includegraphics[width=0.40\linewidth]{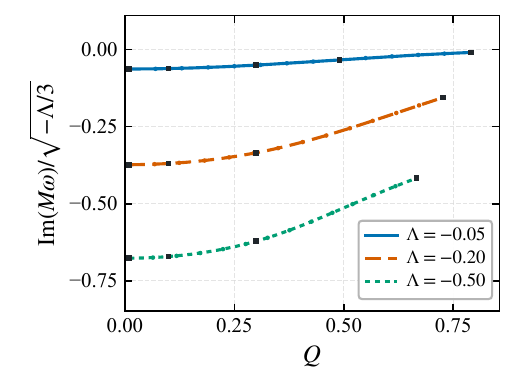}\label{fig:l1-minus-b}}
\\[2pt]
\subfigure[\enspace EM-led mode]{\includegraphics[width=0.40\linewidth]{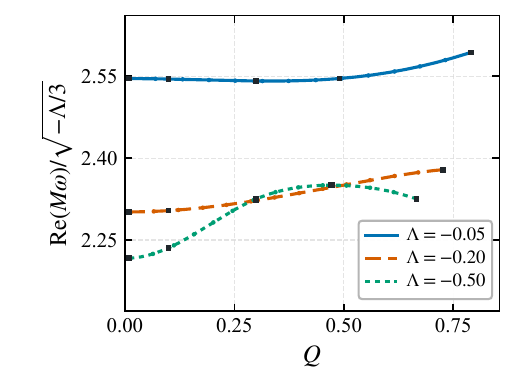}\label{fig:l1-minus-c}}
\hfill
\subfigure[\enspace EM-led mode]{\includegraphics[width=0.40\linewidth]{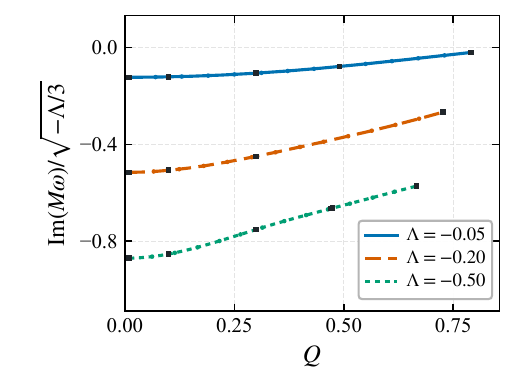}\label{fig:l1-minus-d}}
\caption{Scaled dipole frequencies with fixed $M=1$, $\zeta=-1$ and
$\Lambda=-0.05,-0.20,-0.50$. Normalizations and method symbols
are the same as in Fig.~\ref{fig:l1-plus}.}
\label{fig:l1-minus}
\end{figure}

The charge dependence of the dipole frequencies is shown in
Figs.~\ref{fig:l1-plus} and \ref{fig:l1-minus} for positive and
negative coupling, respectively.
The frequencies are divided by $\sqrt{-\Lambda/3}$ to compare
different curvatures after removing the overall AdS frequency scale.
Lines and circles are obtained by the PS method, and
black squares by DI.
For $\zeta=1$, the charge intervals are $0.01$--$0.59$,
$0.01$--$0.494$ and $0.01$--$0.36$ in order of increasing
$|\Lambda|$, while  for $\zeta=-1$, they are extended  to $0.79$, $0.726$
and $0.666$, respectively.
The effect of the coupling sign is small at weak charge and
it becomes more pronounced as $Q$ increases.
For $\zeta=1$, the scalar-led real frequency in
Fig.~\ref{fig:l1-plus-a} increases with increasing  $Q$ at
$\Lambda=-0.05,-0.20$, but develops a shallow maximum near
$Q=0.312$ at $\Lambda=-0.50$.
Its damping rate in Fig.~\ref{fig:l1-plus-b} decreases throughout all three intervals, so the
scalar perturbation becomes longer lived even after its real frequency
has passed through the maximum.
At a fixed charge, stronger curvature lowers the scaled real frequency
and increases the scaled damping rate.

The electromagnetic (EM)-led mode has a stronger and nonmonotonic charge
dependence, as shown in Figs.~\ref{fig:l1-plus-c} and \ref{fig:l1-plus-d}.
For $\Lambda=-0.05$, its real frequency first decreases to a minimum
near $Q=0.390$, then increases before decreasing again at larger
charge.
The first minimum moves to $Q\simeq0.310$ for $\Lambda=-0.20$,
where the subsequent increase is more pronounced, while for
$\Lambda=-0.50$ the real frequency decreases throughout the displayed
interval.
The damping rate first increases and reaches a local maximum near
$Q=0.484,0.378,0.304$ for the three curvatures, respectively, before
decreasing.
The damping maxima therefore occur at different charges from the
real-frequency extrema.
For example, at $\Lambda=-0.20$ the oscillation frequency increases
between $Q\simeq0.310$ and $0.378$, while the decay time continues
to decrease.
At the largest charges, the damping rates for $\Lambda=-0.05,-0.20$
increase again.
Thus the oscillation frequency and decay time need not vary
in the same direction as the magnetic charge increases.

For $\zeta=-1$, the SC-led real frequency increases monotonically
with charge, with a steeper dependence at stronger curvature, whereas
its damping rate decreases, as shown in Figs.~\ref{fig:l1-minus-a}
and \ref{fig:l1-minus-b}.
The EM-led real frequency in Fig.~\ref{fig:l1-minus-c}
has a shallow minimum near
$Q=0.332$ for $\Lambda=-0.05$, increases throughout the
$\Lambda=-0.20$ interval, and has a broad maximum near $Q=0.472$
for $\Lambda=-0.50$.
Its damping rate in Fig.~\ref{fig:l1-minus-d} decreases throughout all three intervals.
Beyond the maximum at strong curvature, the electromagnetic
perturbation therefore oscillates more slowly and survives longer.
The opposite charge dependence for the two signs of the coupling
is particularly clear over $Q=0.01$--$0.30$ at $\Lambda=-0.50$.
For $\zeta=1$, the EM-led real frequency decreases
and the damping rate increases with charge, whereas the opposite
behavior occurs for $\zeta=-1$. The SC-led real frequency
varies much less for both coupling signs at all three curvatures.
These charge variations contain the effects of both the changing
geometry and the electromagnetic coefficients and  the comparison with
$\zeta=0$ in Sec.~\ref{subsec:nonlinear-response} separates the
contribution of the nonlinear interaction.

Table~\ref{tab:dipole-turning-brackets} gives the charge intervals
containing the first EM-led extrema for $\zeta=1$.
As $|\Lambda|$ increases, the real-frequency minimum and damping
maximum shift towards smaller charges.
For $\Lambda=-0.50$, the real frequency decreases monotonically
along the displayed branch.
The charge intervals depend on the allowed background domain,
with only the oscillatory branch shown for $\zeta=1$
and $\Lambda=-0.50$.

\begin{table}[tbp]

\caption{Grid brackets for the first real-frequency minimum and damping
maximum of the $\zeta=1$ EM-led dipole.}
\label{tab:dipole-turning-brackets}
\centering
\begin{ruledtabular}
\begin{tabular}{ccc}
$\Lambda$ & $Q$ at the first $\omega_R$ minimum &
$Q$ at the first $\gamma$ maximum\\
\hline
$-0.05$ & $[0.388,0.392]$ & $[0.482,0.486]$\\
$-0.20$ & $[0.308,0.312]$ & $[0.376,0.380]$\\
$-0.50$ & none on $[0.01,0.36]$ & $[0.302,0.306]$\\
\end{tabular}
\end{ruledtabular}
\end{table}

\subsection{Nonlinear coupling and weak-charge behavior}
\label{subsec:nonlinear-response}

To examine the NED contribution, we compare the dipole frequencies for $\zeta=\pm1$ with those for $\zeta=0$ at fixed $(M,Q,\Lambda)$.
Since the background dilaton profile is independent of $\zeta$ at
fixed mass and charge, its asymptotic scalar coefficient remains unchanged
in this comparison.
The $\zeta=0$ solution represents a charged dilatonic black hole.
Denoting the corresponding frequency by $\omega_0$, the nonlinear
frequency and damping shifts are defined as
\begin{equation}
 \Delta_R^\zeta=100\left[\frac{\operatorname{Re}\omega_\zeta}
 {\operatorname{Re}\omega_0}-1\right],\qquad
 \Delta_\gamma^\zeta=100\left[\frac{\gamma_\zeta}{\gamma_0}-1\right],
 \qquad \gamma_\zeta=-\operatorname{Im}\omega_\zeta.
 \label{eq:nonlinear-fractional-shift}
\end{equation}
The results at $Q=0.30$ are given in
Table~\ref{tab:nonlinear-baseline}.
For $\zeta=1$, the reduction of the EM-led real
frequency becomes more pronounced as $|\Lambda|$ increases.
At $|\Lambda|=0.50$, its shift reaches about $28\%$, compared
with less than $1\%$ for the SC-led real frequency. The
electromagnetic damping also increases more strongly than the
scalar damping.
The nonlinear interaction therefore reduces the electromagnetic
oscillation frequency and shortens its decay time much more strongly
than for the SC-led mode at this charge.
For negative coupling, the EM-led real frequency instead
increases and its damping rate decreases at all three curvatures.
The EM-led mode is therefore much more sensitive to the NED
interaction than the SC-led mode.

\begin{table}[H]

\caption{Dipole QNM frequencies at $Q=0.30$ with the nonlinear coupling removed,
and fractional shifts when it is restored. All backgrounds use $M=1$
and the same choice of perturbative boundary sources.
The shifts are defined in \eqref{eq:nonlinear-fractional-shift} and  signs on
$\Delta_\gamma$ refer to stronger or weaker damping.}
\label{tab:nonlinear-baseline}
\centering
\begingroup\small\setlength{\tabcolsep}{3.5pt}
\begin{ruledtabular}
\begin{tabular}{clccccc}
$\Lambda$ & mode & $M\omega_0$ & $\Delta_R^{+}$ (\%) &
$\Delta_\gamma^{+}$ (\%) & $\Delta_R^{-}$ (\%) & $\Delta_\gamma^{-}$ (\%)\\
\hline
$-0.05$ & SC & $0.349689028-0.006685531\,\I$ & $-0.06$ & $+1.90$ & $+0.07$ & $-2.04$\\
$-0.05$ & EM & $0.325849360-0.016430614\,\I$ & $-0.58$ & $+24.88$ & $+0.69$ & $-17.04$\\
$-0.20$ & SC & $0.646731442-0.088085513\,\I$ & $-0.21$ & $+1.26$ & $+0.30$ & $-1.82$\\
$-0.20$ & EM & $0.585328736-0.135321308\,\I$ & $-2.11$ & $+29.89$ & $+2.45$ & $-14.27$\\
$-0.50$ & SC & $1.004922236-0.258032818\,\I$ & $-0.55$ & $+1.15$ & $+1.30$ & $-1.63$\\
$-0.50$ & EM & $0.885744859-0.358846748\,\I$ & $-28.14$ & $+34.08$ & $+7.17$ & $-14.49$\\
\end{tabular}
\end{ruledtabular}
\endgroup
\end{table}

The different charge orders of the NED terms can be seen
directly from the radial equations.
At fixed $r$ and negative $\Lambda$, the weak-charge expansion
gives
\begin{align}
 \bar\phi&=\frac{Q^2}{2r}+O(Q^6),\qquad
 \mathcal C=10+O(Q^2),\qquad
 \mathcal D=r^4-20\zeta Q^2+O(Q^4),\nonumber\\
 \frac{\mathcal H}{\mathcal D}
 &=1-\frac{40\zeta Q^2}{r^4}+O(Q^4),\qquad
 A(\zeta,Q,\Lambda)-A(0,Q,\Lambda)=-\frac{2\zeta Q^4}{r^6}.
 \label{eq:weak-charge-operator}
\end{align}
\noindent The electromagnetic angular coefficient contains
$\mathcal H/\mathcal D$ and its derivative coefficient also
depends on $\mathcal D$.
Consequently, the NED correction enters at
$O(\zeta Q^2)$, although the explicit correction to the metric starts
at $O(\zeta Q^4)$.
The scalar diagonal coefficient first depends on the nonlinear
coupling at $O(\zeta Q^4)$.
Since the leading scalar--electromagnetic coupling is $O(Q)$ in
each equation, its feedback into a nondegenerate SC-led mode
adds two powers of $Q$ to the electromagnetic correction.
For nondegenerate dipole frequencies at fixed curvature, the leading
nonlinear shifts are therefore expected to satisfy
\begin{equation}
 \delta\omega_{\rm SC}^\zeta=O(\zeta Q^4),\qquad
 \delta\omega_{\rm EM}^\zeta=O(\zeta Q^2),\qquad
 \delta\omega_b^\zeta=\omega_b(\zeta,Q,\Lambda)-\omega_b(0,Q,\Lambda).
 \label{eq:weak-charge-frequency-hierarchy}
\end{equation}

\begin{table}[H]

\caption{Weak-charge nonlinear dipole shifts at $M=1$ and $\Lambda=-0.05$.
Here, we have  $\delta\omega_b^\zeta=\omega_b(\zeta,Q,\Lambda)-\omega_b(0,Q,\Lambda)$.
The nearly constant scaled magnitudes test the hierarchy in
\eqref{eq:weak-charge-frequency-hierarchy}. Superscripts $+$ and $-$
denote $\zeta=1$ and $-1$.}
\label{tab:weak-charge-hierarchy}
\centering
\begin{ruledtabular}
\begin{tabular}{ccccc}
$Q$ & $|M\delta\omega_{\rm SC}^{+}|/Q^4$ &
$|M\delta\omega_{\rm SC}^{-}|/Q^4$ &
$|M\delta\omega_{\rm EM}^{+}|/Q^2$ &
$|M\delta\omega_{\rm EM}^{-}|/Q^2$\\
\hline
$0.02$ & $0.050592$ & $0.050631$ & $0.042198$ & $0.042163$\\
$0.04$ & $0.050116$ & $0.050274$ & $0.042253$ & $0.042114$\\
$0.08$ & $0.048309$ & $0.048895$ & $0.042491$ & $0.041925$\\
\end{tabular}
\end{ruledtabular}
\end{table}

Table~\ref{tab:weak-charge-hierarchy} confirms these charge orders
for both signs of $\zeta$ at $\Lambda=-0.05$.
Over $Q=0.02$--$0.08$, the nearly constant values of
$|\delta\omega_{\rm SC}^\zeta|/Q^4$ and
$|\delta\omega_{\rm EM}^\zeta|/Q^2$ agree with the leading
powers of four and two. At larger charge, the scalar shift
divided by $Q^4$ begins to vary more strongly, indicating that
higher-order terms can no longer be neglected.

\FloatBarrier

\subsection{Dipole trajectories in the complex plane}
\label{subsec:trajectories}

\begin{figure}[tbp]
\centering
\subfigure[\enspace SC-led mode, $\zeta=1$]{\includegraphics[width=0.45\linewidth]{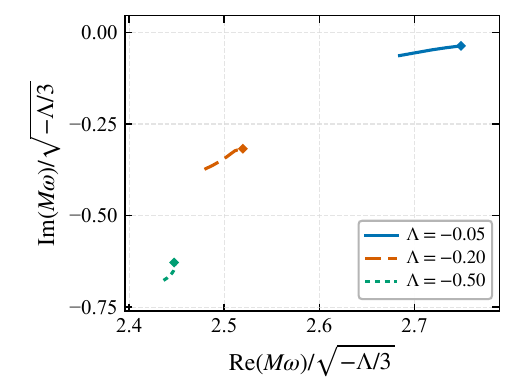}\label{fig:l1-complex-a}}
\hfill
\subfigure[\enspace SC-led mode, $\zeta=-1$]{\includegraphics[width=0.45\linewidth]{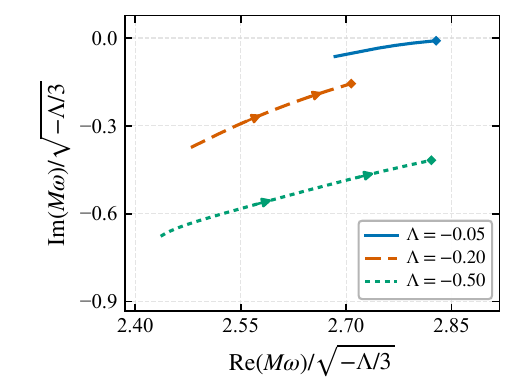}\label{fig:l1-complex-b}}
\\[2pt]
\subfigure[\enspace EM-led mode, $\zeta=1$]{\includegraphics[width=0.45\linewidth]{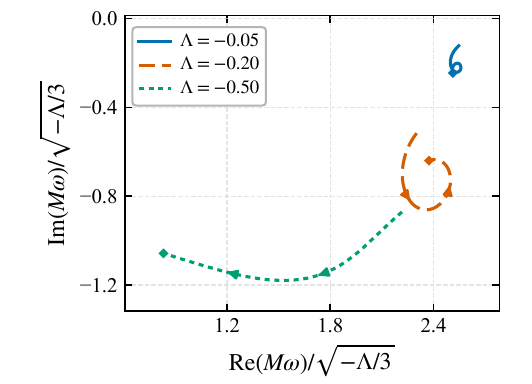}\label{fig:l1-complex-c}}
\hfill
\subfigure[\enspace EM-led mode, $\zeta=-1$]{\includegraphics[width=0.45\linewidth]{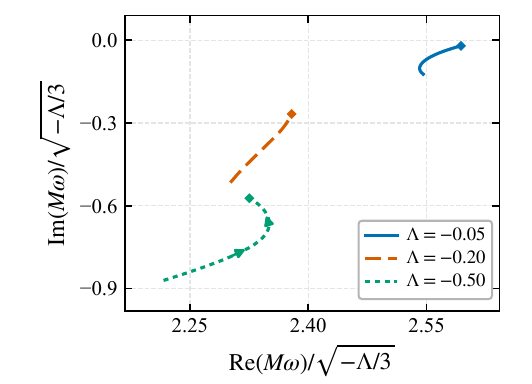}\label{fig:l1-complex-d}}
\caption{Dipole trajectories in the scaled complex-frequency plane
with fixed $M=1$ and $\Lambda=-0.05,-0.20,-0.50$.
Diamonds mark the largest charge on each curve, and arrows indicate
increasing $Q$.}
\label{fig:l1-complex}
\end{figure}

Fig.~\ref{fig:l1-complex} displays the dipole trajectories in the
complex-frequency plane, with both frequency components divided by
$\sqrt{-\Lambda/3}$.
An interesting feature of the EM-led mode for $\zeta=1$ is the
loop-like trajectory at $\Lambda=-0.05$ and the turning trajectory
at $\Lambda=-0.20$, as shown in Fig.~\ref{fig:l1-complex-c}.
These features arise because the oscillation frequency and damping
rate reach their extrema at different charges.
Consequently, black holes with different magnetic charges may
exhibit the same oscillation frequency but different decay times.
At $\Lambda=-0.50$, the trajectory extends towards much smaller
real frequencies and turns towards the real axis after reaching
its maximum damping, indicating that the perturbation oscillates
more slowly but survives longer as the charge increases further.

A different behavior is found for $\zeta=-1$ in
Fig.~\ref{fig:l1-complex-d}, where the EM-led trajectories move
continuously towards the real axis even when the oscillation
frequency reverses its trend.
Increasing the magnetic charge thus prolongs the lifetime
of the EM-led perturbations.
In contrast, the SC-led trajectories in
Figs.~\ref{fig:l1-complex-a} and \ref{fig:l1-complex-b}
exhibit much simpler patterns.
The richer structure of the EM-led trajectories reflects the
stronger influence of the nonlinear electromagnetic interaction
on these modes.

\FloatBarrier

\subsection{Coupled quadrupole modes}
\label{subsec:l2-survey}
\label{subsec:quadrupole-results}

At the reference background $(\zeta,Q,\Lambda)=(1,0.01,-0.05)$,
the quadrupole frequencies in Table~\ref{tab:frequencies} satisfy
$\operatorname{Re}\omega_{\rm GR}<\operatorname{Re}\omega_{\rm EM}
<\operatorname{Re}\omega_{\rm SC}$ and
$\gamma_{\rm GR}<\gamma_{\rm SC}<\gamma_{\rm EM}$.
Their decay times are approximately $2.12\times10^3M$,
$1.04\times10^3M$ and $6.02\times10^2M$ for the GR, SC and EM
modes, respectively, so the GR-led perturbation lives
about 3.5 times longer than the EM-led perturbation.

\begin{figure}[tbp]
\centering
\subfigure[\enspace GR-led mode]{\includegraphics[width=0.48\linewidth]{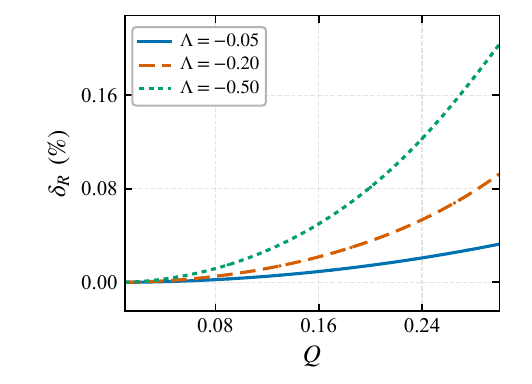}\label{fig:l2-q-plus-a}}
\hfill
\subfigure[\enspace GR-led mode]{\includegraphics[width=0.48\linewidth]{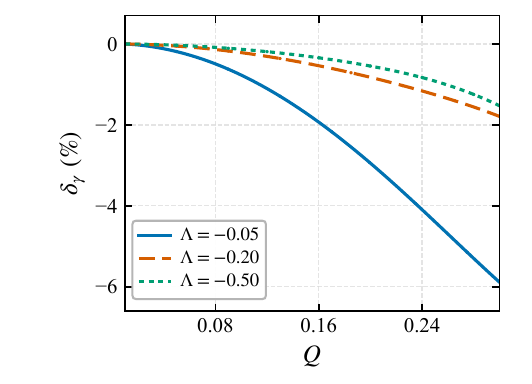}\label{fig:l2-q-plus-b}}
\\[2pt]
\subfigure[\enspace SC-led mode]{\includegraphics[width=0.48\linewidth]{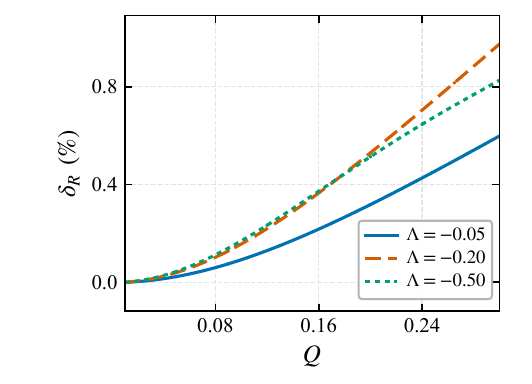}\label{fig:l2-q-plus-c}}
\hfill
\subfigure[\enspace SC-led mode]{\includegraphics[width=0.48\linewidth]{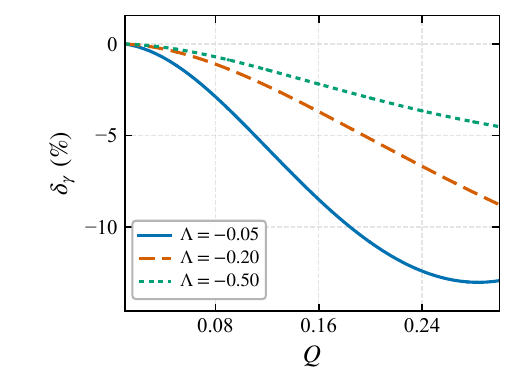}\label{fig:l2-q-plus-d}}
\\[2pt]
\subfigure[\enspace EM-led mode]{\includegraphics[width=0.48\linewidth]{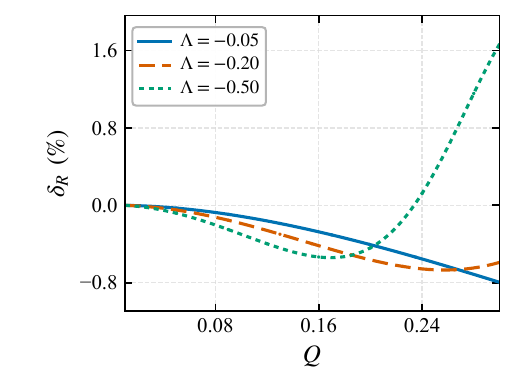}\label{fig:l2-q-plus-e}}
\hfill
\subfigure[\enspace EM-led mode]{\includegraphics[width=0.48\linewidth]{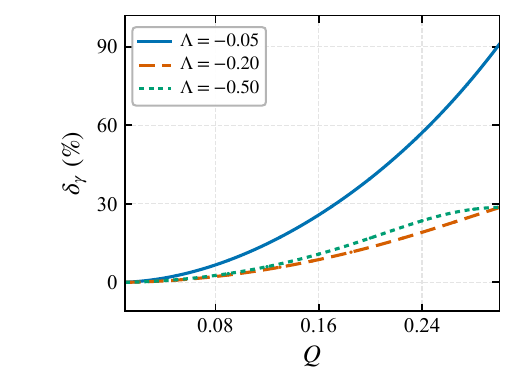}\label{fig:l2-q-plus-f}}
\caption{Charge dependence of the quadrupole modes with fixed $M=1$,
$\zeta=1$ and $\Lambda=-0.05,-0.20,-0.50$.
The left and right columns show $\delta_R$ and $\delta_\gamma$
defined in \eqref{eq:l2-fractional-shift}.}
\label{fig:l2-q-plus}
\end{figure}

\begin{figure}[tbp]
\centering
\subfigure[\enspace GR-led mode]{\includegraphics[width=0.48\linewidth]{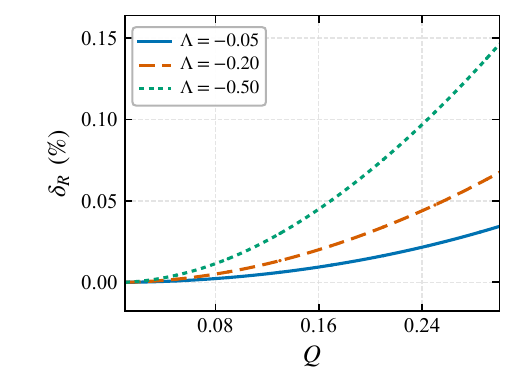}\label{fig:l2-q-minus-a}}
\hfill
\subfigure[\enspace GR-led mode]{\includegraphics[width=0.48\linewidth]{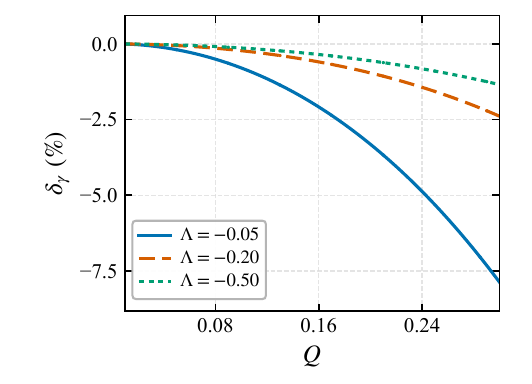}\label{fig:l2-q-minus-b}}
\\[2pt]
\subfigure[\enspace SC-led mode]{\includegraphics[width=0.48\linewidth]{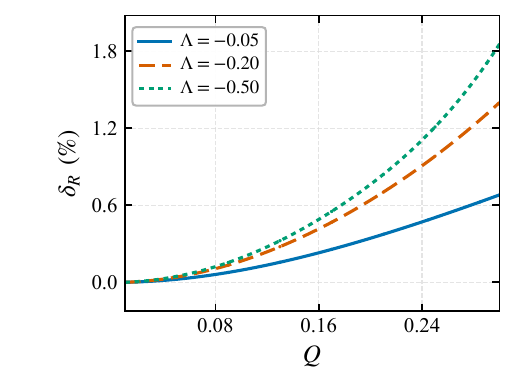}\label{fig:l2-q-minus-c}}
\hfill
\subfigure[\enspace SC-led mode]{\includegraphics[width=0.48\linewidth]{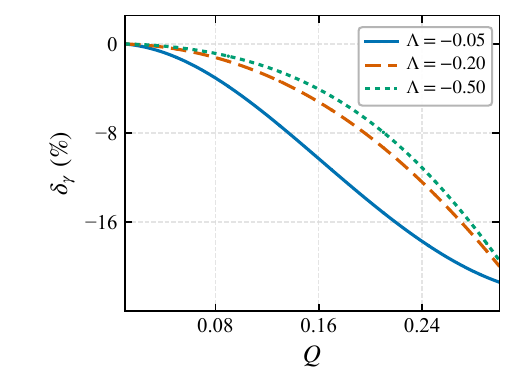}\label{fig:l2-q-minus-d}}
\\[2pt]
\subfigure[\enspace EM-led mode]{\includegraphics[width=0.48\linewidth]{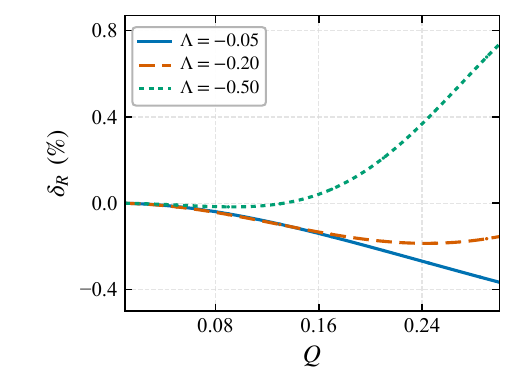}\label{fig:l2-q-minus-e}}
\hfill
\subfigure[\enspace EM-led mode]{\includegraphics[width=0.48\linewidth]{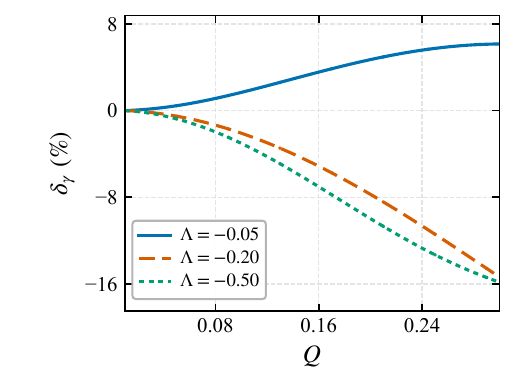}\label{fig:l2-q-minus-f}}
\caption{Charge dependence of the quadrupole modes with fixed $M=1$,
$\zeta=-1$ and $\Lambda=-0.05,-0.20,-0.50$.
Panels and normalizations follow Fig.~\ref{fig:l2-q-plus}.}
\label{fig:l2-q-minus}
\end{figure}

Figs.~\ref{fig:l2-q-plus} and \ref{fig:l2-q-minus} show the charge
dependence of the three quadrupole modes over $Q=0.01$--$0.30$.
Lines and circles are obtained by direct integration.
The SC- and EM-led results occupy the middle and
lower rows, respectively, while the upper row shows the
GR-led mode.
The fractional shifts relative to $Q=0.01$ are defined as
\begin{equation}
 \delta_R=100\left[\frac{\operatorname{Re}\omega(Q)}
 {\operatorname{Re}\omega(0.01)}-1\right],\qquad
 \delta_\gamma=100\left[\frac{\gamma(Q)}{\gamma(0.01)}-1\right],
 \quad \gamma=-\operatorname{Im}\omega>0,
 \label{eq:l2-fractional-shift}
\end{equation}
\noindent where each difference is evaluated at fixed $\zeta$, $\Lambda$ and
mode branch.
The GR-led mode in Figs.~\ref{fig:l2-q-plus-a}, \ref{fig:l2-q-plus-b},
\ref{fig:l2-q-minus-a} and \ref{fig:l2-q-minus-b}
depends weakly on charge. Its real frequency remains nearly
unchanged, while its damping decreases slightly for both coupling
signs. The SC-led real frequency in
Figs.~\ref{fig:l2-q-plus-c} and \ref{fig:l2-q-minus-c}
increases throughout the interval, with a larger variation for
negative coupling.
Its damping, shown in Figs.~\ref{fig:l2-q-plus-d} and
\ref{fig:l2-q-minus-d}, is lower at $Q=0.30$ in all six sequences, so the
SC-led perturbation generally becomes longer lived as the charge
increases over this range.
For $\zeta=1$ and $\Lambda=-0.05$, the damping has a shallow
minimum near $Q=0.285$ before increasing slightly at the final
charges.

The EM-led quadrupole in Figs.~\ref{fig:l2-q-plus-e}, \ref{fig:l2-q-plus-f},
\ref{fig:l2-q-minus-e} and \ref{fig:l2-q-minus-f}
shows a stronger dependence on the nonlinear interaction.
For $\zeta=1$, its damping rate increases with charge at all three
curvatures, so the electromagnetic perturbation decays more rapidly.
Its real frequency is lower at the end of the interval for
$\Lambda=-0.05,-0.20$, whereas for $\Lambda=-0.50$ it first
decreases and then rises above the low-charge value.
For $\zeta=-1$, the damping instead decreases at
$\Lambda=-0.20,-0.50$, and the real frequency has a shallow
minimum at the stronger curvature before increasing.
The larger changes therefore occur in the matter-led branches,
while the direction of the change depends on the mode and the sign
of the coupling.

\begin{figure}[tbp]
\centering
\subfigure[\enspace GR-led mode]{\includegraphics[width=0.48\linewidth]{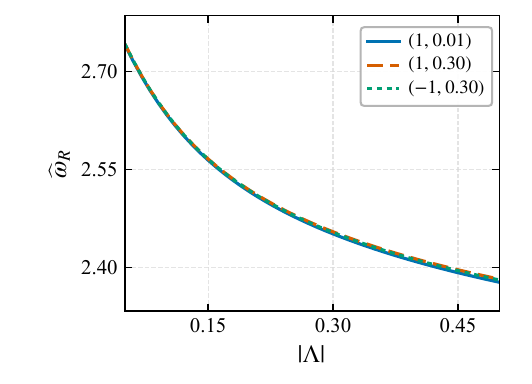}\label{fig:l2-curvature-a}}
\hfill
\subfigure[\enspace GR-led mode]{\includegraphics[width=0.48\linewidth]{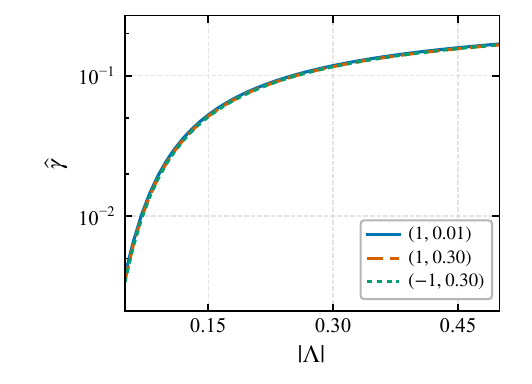}\label{fig:l2-curvature-b}}
\\[2pt]
\subfigure[\enspace SC-led mode]{\includegraphics[width=0.48\linewidth]{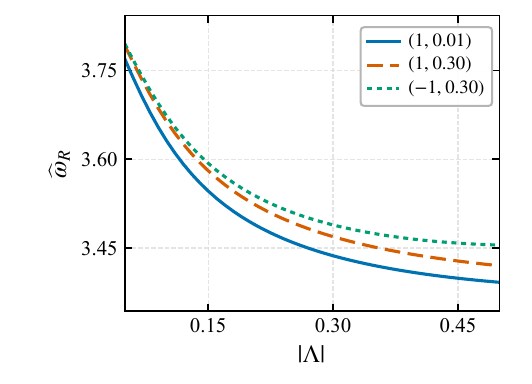}\label{fig:l2-curvature-c}}
\hfill
\subfigure[\enspace SC-led mode]{\includegraphics[width=0.48\linewidth]{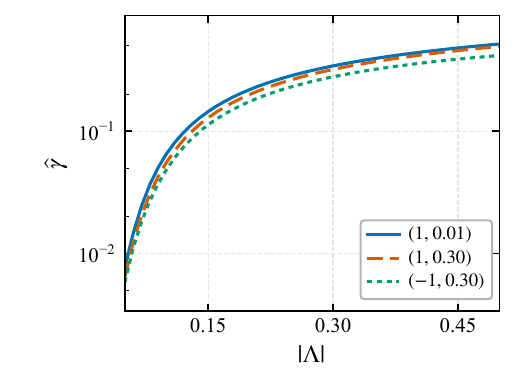}\label{fig:l2-curvature-d}}
\\[2pt]
\subfigure[\enspace EM-led mode]{\includegraphics[width=0.48\linewidth]{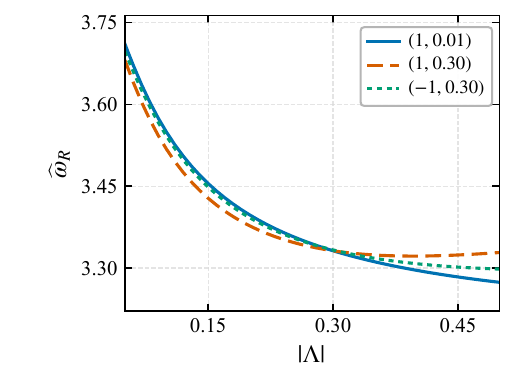}\label{fig:l2-curvature-e}}
\hfill
\subfigure[\enspace EM-led mode]{\includegraphics[width=0.48\linewidth]{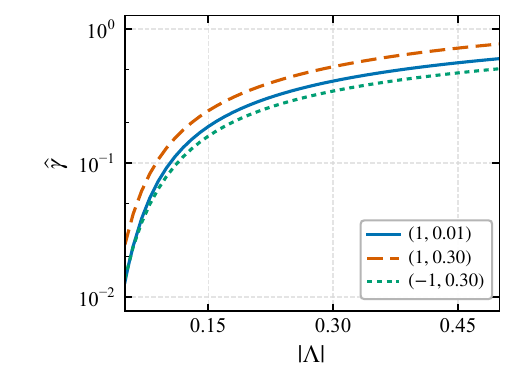}\label{fig:l2-curvature-f}}
\caption{Curvature dependence of the quadrupole modes with fixed $M=1$.
Legend entries denote $(\zeta,Q)=(1,0.01),(1,0.30),(-1,0.30)$.
The left and right columns show $\widehat\omega_R$ and
$\widehat\gamma$, respectively.}
\label{fig:l2-curvature}
\end{figure}

The curvature dependence at $(\zeta,Q)=(1,0.01),(1,0.30),(-1,0.30)$
is shown in Fig.~\ref{fig:l2-curvature} in terms of
$\widehat\omega_R=\operatorname{Re}\omega/\sqrt{|\Lambda|/3}$
and $\widehat\gamma=\gamma/\sqrt{|\Lambda|/3}$.
The legend entries give the pairs $(\zeta,Q)$, and the curves
are obtained by direct integration with PS comparisons
on the same parameter grid.
Over $|\Lambda|=0.05$--$0.50$, all scaled real frequencies are
lower at the strong-curvature end, while the scaled damping rates
are higher.
The positive-coupling EM real frequency at $Q=0.30$ has a shallow
minimum near $|\Lambda|=0.40$.
The GR curves in Figs.~\ref{fig:l2-curvature-a} and
\ref{fig:l2-curvature-b} almost coincide for the three parameter choices,
whereas the SC curves in Figs.~\ref{fig:l2-curvature-c} and
\ref{fig:l2-curvature-d} and the EM curves in
Figs.~\ref{fig:l2-curvature-e} and \ref{fig:l2-curvature-f}
retain a visible dependence on charge and coupling sign.

\subsection{Weak-curvature limit of the dipole modes}
\label{subsec:curvature}

\begin{figure}[tbp]
\centering
\subfigure[\enspace$\zeta=1,\ Q=0.01$]{\includegraphics[width=0.32\linewidth]{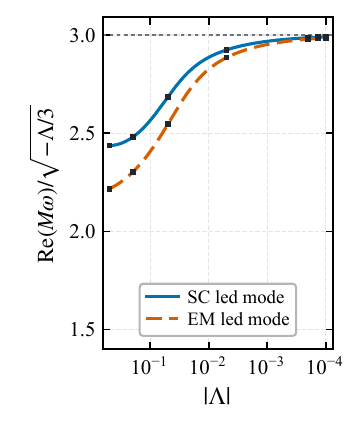}\label{fig:curvature-scaling-a}}
\hfill
\subfigure[\enspace$\zeta=1,\ Q=0.30$]{\includegraphics[width=0.32\linewidth]{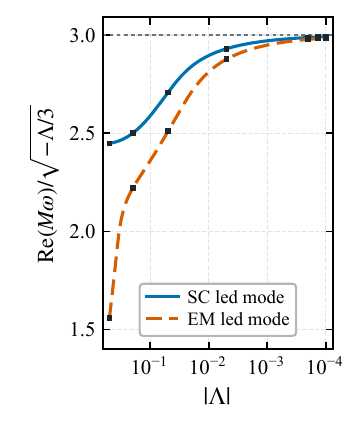}\label{fig:curvature-scaling-b}}
\hfill
\subfigure[\enspace$\zeta=-1,\ Q=0.30$]{\includegraphics[width=0.32\linewidth]{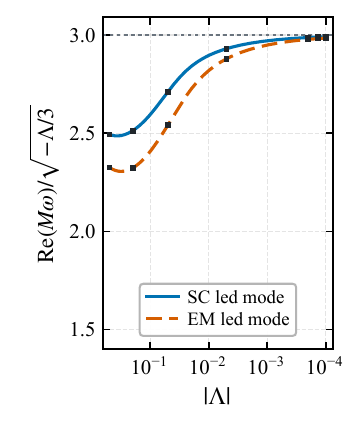}\label{fig:curvature-scaling-c}}
\\[2pt]
\subfigure[\enspace$\zeta=1,\ Q=0.01$]{\includegraphics[width=0.32\linewidth]{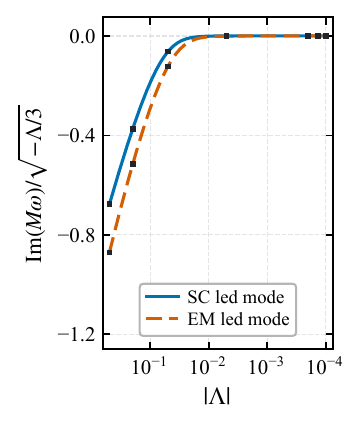}\label{fig:curvature-scaling-d}}
\hfill
\subfigure[\enspace$\zeta=1,\ Q=0.30$]{\includegraphics[width=0.32\linewidth]{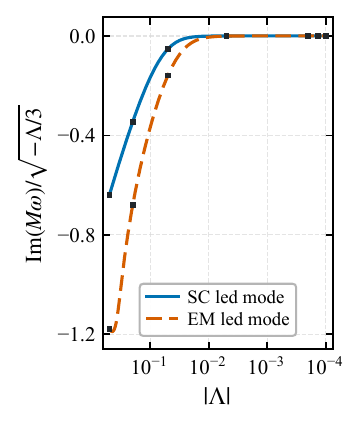}\label{fig:curvature-scaling-e}}
\hfill
\subfigure[\enspace$\zeta=-1,\ Q=0.30$]{\includegraphics[width=0.32\linewidth]{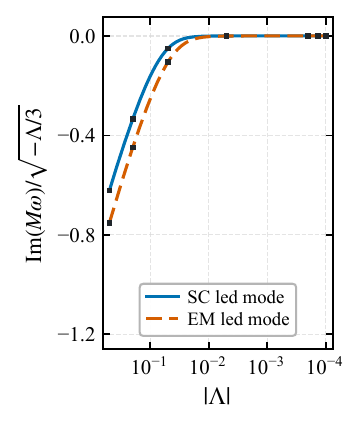}\label{fig:curvature-scaling-f}}
\caption{Curvature dependence of the dipole frequencies with fixed $M=1$.
The upper and lower rows show the real and imaginary parts of
$M\omega/\sqrt{-\Lambda/3}$, respectively.
The gray dashed line indicates the pure AdS real-frequency value $3$.}
\label{fig:curvature-scaling}
\end{figure}

The dipole frequencies are followed from $\Lambda=-0.50$ to
$-0.0001$ at $(\zeta,Q)=(1,0.01),(1,0.30),(-1,0.30)$, as shown
in Fig.~\ref{fig:curvature-scaling}.
Solid blue and dashed orange curves give the SC-led and
EM-led PS frequencies, respectively,
and black squares are obtained by direct integration.
The pronounced dependence of the dipole frequencies on charge
and nonlinear coupling at strong curvature gradually weakens
as $\Lambda$ approaches zero.

As $|\Lambda|$ decreases, the six scaled real frequencies approach
the value 3 from below, and their differences at the three parameter
choices become smaller.
For $|\Lambda|\lesssim0.05$, the scaled imaginary parts also
approach zero from below.
The decrease of $\gamma/\sqrt{-\Lambda/3}$ means that the decay
time grows relative to the oscillation period, so the perturbation
undergoes more oscillations before decaying.
Nevertheless, the final frequencies in Table~\ref{tab:weak-curvature}
still have different damping rates and  the SC-led decay times are
about four to nine times those of the EM-led modes at
the same parameters.
For the scalar field with standard quantization, the lowest-radial-order
normal frequency in pure AdS satisfies $\omega L=\ell+\Delta=3$,
where $\Delta=2$ and $L=\sqrt{-3/\Lambda}$.
The EM-led branch approaches the same real-frequency
scale.
At fixed mass, the horizon becomes small relative to the AdS radius
as $\Lambda\to0^-$, so the approach to normal frequencies with
decreasing damping agrees with the small AdS black hole
limit~\cite{Konoplya:2002zu}.

\FloatBarrier

\begin{table}[H]
\caption{Dipole spectral frequencies at the final continuation point,
$M=1$ and $\Lambda=-0.0001$. The last column denotes 
$\operatorname{Re}(M\omega)/\sqrt{-\Lambda/3}$.}
\label{tab:weak-curvature}
\centering
\begin{ruledtabular}
\begin{tabular}{ccccc}
$\zeta$, $Q$ & mode & $\operatorname{Re}(M\omega)$ & $\operatorname{Im}(M\omega)$ & scaled real part\\
\hline
1, 0.01 & SC & 0.017263258499 & $-1.4422\times10^{-10}$ & 2.99008408\\
1, 0.01 & EM & 0.017234544332 & $-5.7253\times10^{-10}$ & 2.98511064\\
1, 0.30 & SC & 0.017267587457 & $-1.0210\times10^{-10}$ & 2.99083388\\
1, 0.30 & EM & 0.017228980345 & $-8.7429\times10^{-10}$ & 2.98414693\\
-1, 0.30 & SC & 0.017267589372 & $-9.7680\times10^{-11}$ & 2.99083421\\
-1, 0.30 & EM & 0.017229008811 & $-4.8562\times10^{-10}$ & 2.98415186\\
\end{tabular}
\end{ruledtabular}
\end{table}

\noindent\begin{minipage}{\linewidth}
\setlength{\parindent}{1.5em}\indent
As $\Lambda \to 0^-$, the region of $|\Lambda|r^2=O(1)$ moves to
larger radii, making the limit nonuniform in $r$.
The reflecting AdS boundary condition is retained in this limit, 
whereas asymptotically flat QNMs obey an outgoing condition at infinity.
\end{minipage}

\FloatBarrier

\subsection{Connection between phase transition and \texorpdfstring{\ensuremath{\ell=0}}{l=0} dilaton}
\label{subsec:thermo-dilaton-qnm}

It is important to explore connection between phase transition and $\ell=0$ mode dilaton.
For this purpose, we introduce  the $\ell=0$ dilaton  at fixed  $Q_m=0.4$, $\epsilon=-0.5$ and $P=0.6P_c$. The coexistence temperature is given by $T_*=0.0729629$ with horizon radii $r_s=0.951028$ and $r_l=2.575539$. These states have equal temperature and Gibbs free energy. See  Appendix~\ref{app:thermo-dilaton} for summarizing two thermodynamic branches of small black holes (SBH) and large black holes (LBH).

\begin{figure}[tbp]
\centering
\includegraphics[width=0.94\linewidth]{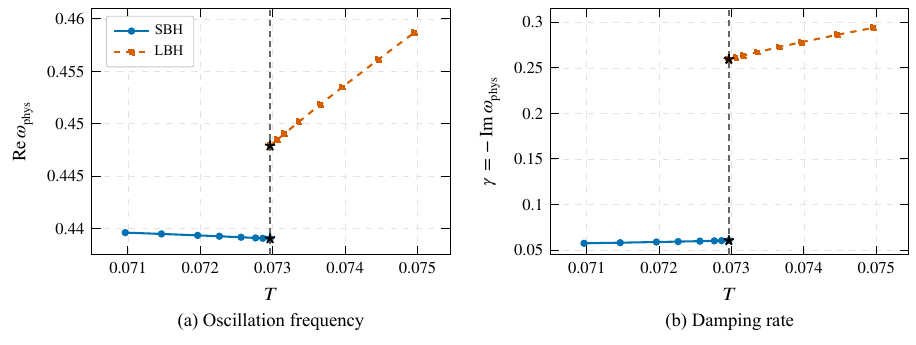}
\caption{$\ell=0$ dilaton QNMs versus temperature for $Q_m=0.4$, $\epsilon=-0.5$ and $P=0.6P_c$ with standard quantization ($a_0=0$). Panel (a) shows the oscillation frequency, while panel (b) indicates the damping rate on the SBH (blue) and LBH (orange) branches. Markers denote computed backgrounds, stars mark coexistence, and the vertical dashed line represents  $T_*$. Frequencies are expressed in physical units for each background.}
\label{fig:dilaton-qnm-temperature}
\end{figure}

Figure~\ref{fig:dilaton-qnm-temperature} shows the monopole frequencies on the preferred branches, including coexistence, with standard boundary condition $a_0=0$. Temperature and frequency are expressed in terms of physical units, accounting for the varying mass along the fixed-pressure path. We note that DI and PS calculations agree for all sampled backgrounds.

As the temperature rises toward coexistence along the SBH branch, $\operatorname{Re}\omega_{\rm phys}$ decreases slightly while the damping rate $\gamma=-\operatorname{Im}\omega_{\rm phys}$ increases. Along the LBH branch, however,  both quantities increase with temperature. At coexistence,  two real frequencies differ by $0.00885$, whereas the damping rates differ by $0.19876$. The LBH branch is damped about 4.28 times faster than the SBH branch. Compared with the massless external scalar on the same thermodynamic backgrounds~\cite{Chen:2026mxq}, the dilaton has a smaller absolute frequency separation but a larger damping-rate ratio than the massless scalar.

\FloatBarrier

\section{Conclusions and discussion}
\label{sec:conclusions}

In this paper, we have obtained  the Type-II QNMs of magnetically
charged dEH-AdS black holes.
The coupled perturbation equations for $\ell=0,1,2$ modes were derived
with the contributions of the cosmological dilaton potential retained
and their QNM frequencies were computed  by DI and
Chebyshev PS methods.
We were aware that  two calculations agree with each other under ingoing horizon
conditions and fixed perturbative sources at the AdS boundary.

Comparing with the charged dilatonic solution (GMGHS) at $\zeta=0$, we
find that the nonlinear correction enters the EM-led
dipole frequency at order $Q^2$, whereas the SC-led correction
starts at order $Q^4$ for nondegenerate modes at fixed curvature.
This difference follows from the charge orders of the NED
terms in the coupled radial equations.

The EM-led dipole is particularly sensitive to the NED interaction.
At $Q=0.30$, stronger AdS curvature enhances the reduction of
its real frequency for positive coupling, whereas negative
coupling increases its real frequency and reduces its damping.

The real-frequency and damping extrema occur at different charges
and shift as the cosmological constant $\Lambda$ changes, so the oscillation
frequency and decay time need not increase or decrease together.

For the quadrupole modes, the GR-led real frequency shows
a much weaker charge dependence than those of the SC- and
EM-led modes.
As the curvature weakens, the scaled dipole real frequencies
approach the AdS normal-mode value and their damping decreases,
corresponding to longer-lived perturbations of black holes that
are small relative to the AdS radius.

All the computed Type-II QNMs with $\ell=0,1,2$ have
negative imaginary parts, providing evidence for the
linear stability of nonextremal dEH-AdS black holes.


Furthermore, we have also examined the $\ell=0$ dilaton QNMs along the SBH and LBH  branches for a first-order phase transition at $P=0.6P_c$. The real frequency decreases with increasing temperature on the SBH branch, whereas it increases on the LBH branch. At coexistence, the two modes have similar oscillation frequencies but markedly different damping rates, with the LBH branch  decaying about 4.28 times faster than the SBH branch. These results are considered as a better test than the massless external scalar analysis of Ref.~\cite{Chen:2026mxq} to  the underlying same black holes.

These results show that the dynamical dilaton QNMs
can distinguish the small- and large-black-hole phases
through their markedly different damping rates.

\FloatBarrier

\par\vspace{1cm}
\noindent\begin{minipage}{\linewidth}
{\bf Acknowledgments}\par\medskip
We thank Sheng-Yuan Li for his contributions to this work.
F. W. Shu was supported by the National Natural Science Foundation of
China (Grant No.~12375049), Key Program of the Natural Science Foundation
of Jiangxi Province under Grant No.~20232ACB201008, and the Ganpo High-Level
Innovative Talent Program with no. gpyc20240060. D. C. Zou was supported by the National Natural
Science Foundation of China (NNSFC) (Grant No.~12365009) and Jiangxi
Provincial Natural Science Foundation (Grant No.~20262BAC240347).\par
\end{minipage}
\par\vspace{1cm}

\appendix
\section{Quadrupole constraint and AdS boundary solutions}
\label{app:operator}

\subsection{Quadrupole algebraic constraint}
\label{ Quadrupole algebraic constraint}

The coefficients of the algebraic constraint
\eqref{eq:l2-constraint} are given by
\begin{align}
C_H={}&rA\left(4r^6e^{2\bar\phi}+2r^4Q^2-2\zeta Q^4{\cal C}
+r^8e^{2\bar\phi}\bar V\right)+3r^8Be^{2\bar\phi}A',
\label{eq:l2-constraint-H}\\
C_R={}&-\I Be^{2\bar\phi}\left(r^9A'{\cal G}+2r^8\omega^2\right),
\label{eq:l2-constraint-R}\\
C_K={}&rA\left(-4r^6e^{2\bar\phi}-4r^4Q^2+8\zeta Q^4{\cal C}\right)
+\frac{r^9Be^{2\bar\phi}(A')^2}{2A}\nonumber\\
&+r^8e^{2\bar\phi}(2r\omega^2-BA'),
\label{eq:l2-constraint-K}\\
C_u={}&24rAQ{\cal D},\qquad
C_{u'}=8r^2ABQ{\cal D},
\label{eq:l2-constraint-u}\\
C_\phi={}&-2r^8Be^{2\bar\phi}A'\bar\phi'
-4A\left[r^7Be^{2\bar\phi}\bar\phi'+r^4Q^2
+3\zeta Q^4(e^{4\bar\phi}-1)\right]\nonumber\\
&+Ar^8e^{2\bar\phi}V_1,\qquad
C_{\phi'}=-4r^8ABe^{2\bar\phi}\bar\phi'.
\label{eq:l2-constraint-phi}
\end{align}
The terms containing $\bar V$ and $V_1$ arise from the cosmological
dilaton potential.
For $C_H\ne0$, this relation determines $H_0$ and leaves the
six-component state in \eqref{eq:l2-physical-state}.
At the reference parameters $(\zeta,Q,\Lambda)=(1,0.01,-0.05)$,
the factorization
\begin{equation}
 C_H=r^7Ae^{2\bar\phi}\widehat C_H,\qquad
 \widehat C_H=4+\frac{2e^{-2\bar\phi}Q^2}{r^2}
 -\frac{2\zeta e^{-2\bar\phi}Q^4{\cal C}}{r^6}+r^2\bar V
 +\frac{3rBA'}{A}.
 \label{eq:constraint-domain}
\end{equation}
has $\widehat C_H>0$ throughout the exterior.  After substituting
the rational $y$-coordinate background, the denominator is positive
and the numerator has positive coefficients in $y-9/5$, for
$y_h>9/5$.
The horizon zero of $A$ is treated by the ingoing expansion.
For the numerical evolution, $H_0$ is reconstructed as
$H_0=-(C_RR_h+C_KK_p+C_u u_4+C_{u'}u_4'
+C_\phi\phi_1+C_{\phi'}\phi_1')/C_H$
and it is substituted into the remaining six equations.

\subsection{Source-free asymptotic solutions}
\label{app:boundary-subspaces}

For $s=1/r$, the scaled quadrupole state is defined by
\begin{equation}
 W_2=(K_p,rR_h,\phi_1,r\phi_1',u_4,ru_4')^{\mathsf T},
 \qquad s\frac{\mathrm dW_2}{\mathrm ds}
       =(\mathsf R_0+O(s))W_2.
 \label{eq:l2-boundary-state}
\end{equation}
Substitution of the exact background gives
$\mathsf R_0^2=\mathsf R_0$, with three slow exponents (0) and
three fast exponents (1).
For the monopole and dipole, these exponents occur once and twice,
respectively.
For the quadrupole, we define
\begin{align}
 d_\Lambda&=\Lambda Q^4-24,\qquad
 a_\omega=\frac{36(\Lambda+\omega^2)}{\Lambda d_\Lambda},\qquad
 b_\omega=\frac{12\I(\Lambda+\omega^2)}{d_\Lambda},\nonumber\\
 c_\phi&=\frac{4\Lambda Q^2}{d_\Lambda},\qquad
 c_u=\frac{16\Lambda Q}{d_\Lambda}.
 \label{eq:l2-boundary-coefficients}
\end{align}
Three independent rows of $I-\mathsf R_0$ indicate  the source matrix as 
\begin{equation}
 \mathsf B_{\rm src}=
 \begin{pmatrix}
 1-a_\omega&-b_\omega&0&-c_\phi&0&-c_u\\
 0&0&1&1&0&0\\
 0&0&0&0&1&1
 \end{pmatrix}.
 \label{eq:l2-source-matrix}
\end{equation}
Writing $W_2=\mathbf c_0+s\mathbf c_1+\cdots$, the radial
equations require $\mathsf R_0\mathbf c_0=0$.
The source-free condition of $\mathsf B_{\rm src}\mathbf c_0=0$
then gives $\mathbf c_0=0$, which is equivalent to
$k_0=a_0=b_0=0$ in Sec.~\ref{subsec:boundary}.
The retained leading vectors satisfy
\begin{equation}
 (\mathsf R_0-I)\mathbf c_1=0,
 \qquad \mathsf B_{\rm src}\mathbf c_1=0,
 \label{eq:l2-fast-space}
\end{equation}
and are normalized by their independent $K_p$, $\phi_1$ and
$u_4$ amplitudes.
The generic leading powers are given by $K_p,H_0,\phi_1,u_4=O(r^{-1})$
and $R_h=O(r^{-2})$, while the constraint gives
$H_0=-K_p+O(r^{-2})$.
An induced component can decay faster if its leading coefficient
vanishes.
Thus, three independent source conditions determine the boundary
subspace, and the coupled equations fix the remaining components
and their derivatives.

\FloatBarrier
\section{Thermodynamic branches and \texorpdfstring{\ensuremath{\ell=0}}{l=0} dilaton modes}
\label{app:thermo-dilaton}

\subsection{Coexistence thermodynamic backgrounds}

\begin{figure}[tbp]
\centering
\includegraphics[width=0.60\linewidth]{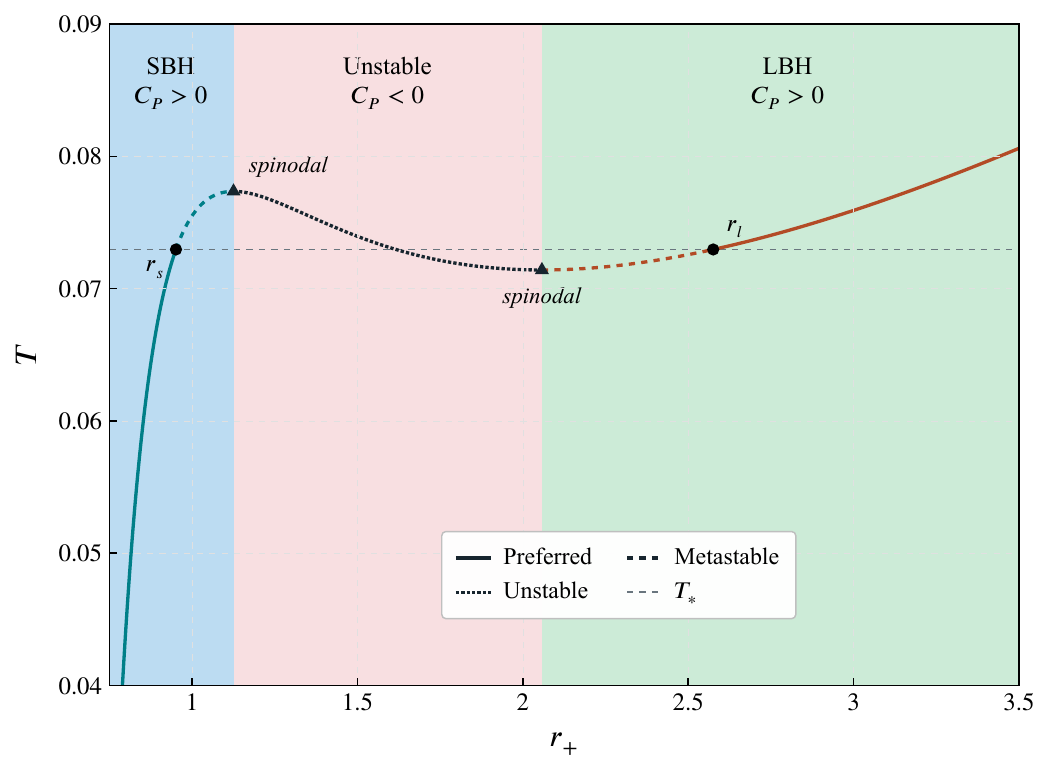}
\caption{Temperature versus horizon radius near coexistence at $P=0.6P_c$, $Q_m=0.4$ and $\epsilon=-0.5$. Blue and green shading indicate locally stable SBH and LBH regions and  pink marks the unstable interval. Solid curves show globally preferred branches, while dashed curves their metastable continuations. Circles mark coexistence and  triangles denote the Davies (spinodal) points.}
\label{fig:thermo-branches}
\end{figure}

For comparison with Ref.~\cite{Chen:2026mxq}, we introduce  the extended-phase-space pressure $P=-\Lambda_{\rm phys}/(8\pi)$ and fix  the  magnetic charge $Q_m=0.4$, EH coupling $\epsilon=-0.5$, and pressure $P=0.6P_c$  with $P_c=0.0138490679$. The horizon relation and thermodynamic quantities are given by 
\begin{align}
K(r_+) &= r_+^2+Q_m^2-\frac{2\epsilon Q_m^4}{r_+^4}+\frac{8\pi P r_+^4}{3}, \nonumber\\
\Delta(r_+) &= \sqrt{K(r_+)^2-Q_m^4},\qquad M=\frac{\Delta}{2r_+}, \nonumber\\
T &= \frac{r_+ K K'-\Delta^2}{4\pi r_+^3\Delta},\qquad
S=\pi r_+^2,\qquad G=M-TS .
\label{eq:thermo-dilaton-functions}
\end{align}
Here $r_+$ is the outer horizon radius, $M$ is the black hole mass, $K$ and $\Delta$ are auxiliary functions entering the horizon relation, and $K'=\partial K/\partial r_+$ at fixed $P$, $Q_m$, and $\epsilon$. Thermodynamic quantities are the Hawking temperature $T$, entropy $S$, and Gibbs free energy $G$.  We retain the positive mass outer horizon branch with $K>Q_m^2$ and $T>0$. Differentiating $K$ gives
\begin{equation}
K'=2r_++\frac{8\epsilon Q_m^4}{r_+^5}+\frac{32\pi P r_+^3}{3}.
\label{eq:thermo-dilaton-kprime}
\end{equation}
Substituting $K'$ into Eq.~\eqref{eq:thermo-dilaton-functions} leads to the Hawking temperature at fixed $Q_m$ and $\epsilon$:
\begin{equation}
\boxed{T(r_+,P)=\frac{K(r_+)\left(r_+^2-Q_m^2+\dfrac{10\epsilon Q_m^4}{r_+^4}+8\pi P r_+^4\right)+Q_m^4}{4\pi r_+^3\sqrt{K(r_+)^2-Q_m^4}}.}
\label{eq:thermo-dilaton-eos}
\end{equation}
For $P<P_c$, the temperature curve has two extrema, which are Davies (spinodal) points where the heat capacity $C_P=2\pi r_+T/(\partial T/\partial r_+)_P$ diverges. The first-order coexistence point is fixed by equal temperatures and Gibbs free energies on the SBH and LBH branches. At $P=0.6P_c$, we find $T_*=0.0729628848$, $r_s=0.9510278173$, and $r_l=2.5755387507$, with $G_*=0.3928404406$. The region between  Davies radii $r_+=1.1250662$ and $2.0575477$ represent the locally unstable segment because its heat capacity is negative. Fig.~\ref{fig:thermo-branches} shows the preferred and metastable portions. The two outer branches have $C_P>0$ and thus, they are locally stable. In addition, their Gibbs energy crossing selects the SBH branch  below $T_*$ and the LBH branch above it, separating local stability from global preference.

\subsection{Frequency branches}

\begin{figure}[tbp]
\centering
\includegraphics[width=0.94\linewidth]{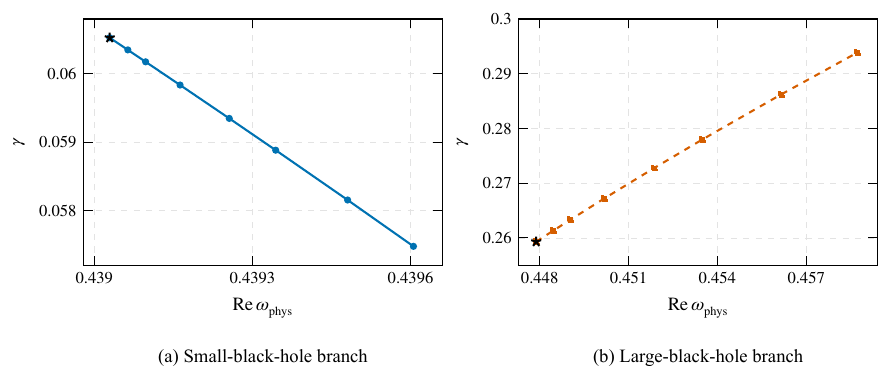}
\caption{QNM frequencies of the $\ell=0$ dilaton  at $Q_m=0.4$, $\epsilon=-0.5$ and $P=0.6P_c$. Panel (a) shows the SBH branch and panel (b) indicates the LBH branch. The axes show $\operatorname{Re}\omega_{\rm phys}$ vs  $\gamma=-\operatorname{Im}\omega_{\rm phys}$. Circular (square) markers denote computed SBH (LBH) backgrounds, whereas  black stars mark the coexistence modes.}
\label{fig:dilaton-qnm-plane}
\end{figure}

The frequencies are given in physical units as  $\omega_{\rm phys}=\widehat\omega/M$ because the mass varies along the fixed-pressure path.
Fig.~\ref{fig:dilaton-qnm-plane} shows the two branches separately. Along the SBH branch, damping increases toward coexistence as the real frequency decreases slightly. On the LBH branch, both the real frequency and damping increase with temperature.
We note that the frequencies at the coexistence endpoints are
\begin{equation}
\omega_s=0.439029169377-0.060525365354\,\I,\qquad
\omega_l=0.447881936650-0.259282140829\,\I .
\label{eq:thermo-dilaton-coexistence-frequencies}
\end{equation}
The frequency separation indicates $|\omega_l-\omega_s|=0.198954$ and the damping-rate ratio is given by $\gamma_l/\gamma_s=4.28386$. The corresponding masses are given by $M_s=0.600159061$ and $M_l=1.913346118$.
On the other hand, the massless test scalar in Ref.~\cite{Chen:2026mxq} gives a larger coexistence-frequency separation ($0.308249$) and a smaller damping-rate ratio ($2.77923$).
Table~\ref{tab:thermo-dilaton-samples} lists representative frequencies on both branches.

\begin{table}[H]
\centering
\caption{QNM frequencies for $\ell=0$  dilaton  on the two thermodynamic branches of SBH and LBH. Frequencies are given in physical units with $\gamma=-\operatorname{Im}\omega_{\rm phys}$.}
\label{tab:thermo-dilaton-samples}
\begin{ruledtabular}
\begin{tabular}{lccccc}
Branch & $T-T_*$ & $r_+$ & $M$ & $\operatorname{Re}\omega_{\rm phys}$ & $\gamma$\\
\hline
SBH & $-0.0020$ & 0.92647678 & 0.58973036 & 0.43960556 & 0.05747502\\
SBH & $-0.0010$ & 0.93785527 & 0.59449371 & 0.43934417 & 0.05888130\\
SBH & $0$ & 0.95102782 & 0.60015906 & 0.43902917 & 0.06052537\\
LBH & $0$ & 2.57553875 & 1.91334612 & 0.44788194 & 0.25928214\\
LBH & $+0.0010$ & 2.73946553 & 2.11439640 & 0.45350411 & 0.27800414\\
LBH & $+0.0020$ & 2.87956865 & 2.29854280 & 0.45873630 & 0.29389208\\
\end{tabular}
\end{ruledtabular}
\end{table}


\FloatBarrier
\bibliographystyle{unsrtnat}
\begingroup
\renewcommand{\baselinestretch}{1.3}\small
\setlength{\bibsep}{3pt plus 1pt minus 1pt}
\interlinepenalty=10000
\bibliography{references}
\endgroup
\end{document}